\documentclass[apj]{emulateapj}
\newcommand{\I}{IGR J}

\usepackage{color}
\usepackage{graphicx}
\usepackage{textcomp}
\usepackage{amssymb}
\usepackage[
      dvipdfmx,    
      colorlinks=true,    
      citecolor=black,
      urlcolor=black,    
      menucolor=black,    
      linkcolor=black,    
      pagecolor=black,    
      bookmarks=true,    
      bookmarksopen=true,    
      hyperfootnotes=false,    
      pdfpagemode=UseOutlines    
]{hyperref}
\shorttitle{Investigating the Optical Counterpart Candidates of \emph{INTEGRAL} Sources}
\shortauthors{\"Ozbey Arabac{\i} et al.}
\begin{document}

\title{Investigating the Optical Counterpart Candidates of Four \emph{INTEGRAL} Sources localized with \emph{Chandra}}
\author{Mehtap \"Ozbey Arabac{\i}\altaffilmark{1}, Emrah Kalemci\altaffilmark{2}, John A. Tomsick\altaffilmark{3}, Jules Halpern\altaffilmark{4}, Arash Bodaghee\altaffilmark{3}, Sylvain Chaty\altaffilmark{5,6}, Jerome Rodriguez\altaffilmark{5}, Farid Rahoui\altaffilmark{7,8}}
\affil{}
\altaffiltext{1}{Middle East Technical University, Department of Physics, Ankara, 06531, TURKEY}
\altaffiltext{2}{Faculty of Engineering and Natural Sciences, Sabanc{\i} University, Orhanl{\i}-Tuzla, \.{I}stanbul, 34596, TURKEY}
\altaffiltext{3}{Space Sciences Laboratory, 7 Gauss Way, University of California, Berkeley, CA 94720-7450, USA}
\altaffiltext{4}{Columbia Astrophysics Laboratory, Columbia University, 550 West 120th Street, New York, NY 10027-6601, USA}
\altaffiltext{5}{AIM (UMR-E 9005 CEA/DSM-CNRS-Universit\'{e} Paris Diderot) Irfu/Service d'Astrophysique, Centre de Saclay, FR-91191 Gif-sur-Yvette Cedex, FRANCE}
\altaffiltext{6}{Institut Universitaire de France, 103, bd Saint-Michel, 75005 Paris, France}
\altaffiltext{7}{Harvard University, Astronomy Department, 60 Garden Street, Cambridge, MA 02138, USA}
\altaffiltext{8}{Harvard-Smithsonian Center for Astrophysics, 60 Garden Street, Cambridge, MA 02138, USA}

\begin{abstract}

We report on the optical spectroscopic follow up observations of the candidate counterparts to four 
\emph{INTEGRAL} sources: IGR J04069$+$5042, IGR J06552$-$1146, IGR J21188$+$4901 and IGR J22014$+$6034. 
The candidate counterparts were determined with \emph{Chandra}, and the optical observations were performed with 1.5-m RTT-150 telescope 
(T\"{U}B\.{I}TAK National Observatory, Antalya, Turkey) and 2.4-m Hiltner Telescope (MDM Observatory, Kitt Peak, Arizona). 
Our spectroscopic results show that one of the two candidates of IGR J04069$+$5042 and the one observed for IGR J06552$-$1146 
could be active late-type stars in RS CVn systems. However, according to the likelihood analysis based on \emph{Chandra} 
and \emph{INTEGRAL}, two optically weaker sources in the \emph{INTEGRAL} error circle of IGR J06552$-$1146 have higher 
probabilities to be the actual counterpart. The candidate counterparts of IGR J21188$+$4901 are
classified as an active M-type star and a late-type star. Among the optical spectra of four candidates of
IGR J22014$+$6034, two show H$\alpha$ emission lines, one is a late-type star and the other is a M type. The likelihood analysis 
favors a candidate with no distinguishing features in the optical spectrum. Two of the candidates classified as M type dwarfs 
are similar to some IGR candidates claimed to be symbiotic stars. However, some of the prominent features of symbiotic systems are 
missing in our spectra, and their NIR colors are not consistent with those expected for giants. We consider the IR colors of all 
IGR candidates claimed to be symbiotic systems and find that low resolution optical spectrum may not be enough for conclusive 
identification.
\end{abstract}             

\keywords{Stars: late-type -- Stars: activity -- Stars: flare -- binaries: symbiotics -- X-rays:individuals -- X-ray:binaries}
    
\section{Introduction}

  The International Gamma-Ray Astrophysics Laboratory ({\emph{INTEGRAL}; \citealt{wink03}) has discovered hundreds 
  of new hard X-ray sources (so called ``IGR'' sources) at energies above 20 keV since its launch on October 2002. From over 700 sources 
  that are listed in the 4$^{th}$ Bird Catalog \citep{bird10}, more than 400 of them are exclusively IGR 
  sources\footnote{See \url{http://irfu.cea.fr/Sap/IGR-Sources/} for an updated list}. {\emph{INTEGRAL}} is very successful in 
  finding new hard X-ray sources thanks to its wide field of view, observing strategy and most importantly unique imaging capability of 
  the ISGRI detector \citep{lebrun03} with a localization accuracy of a few arcmin. However, such localization accuracy 
  is often not enough to identify the correct optical counterparts of these sources. Without multiwavelength information, it may not be 
  possible to understand the physical origin of the X-ray emission and type of the emitting system. To identify a unique optical or 
  infrared counterpart, arcsecond accuracy localization of the X-ray source is required.
 
  Our group observes the positions of IGR sources with soft X-ray telescopes such as {\emph{Chandra}} and {\emph{Swift}},
  not only for identifying the correct counterparts, but also for producing 0.3$-$10 keV spectra that can be used to measure column 
  densities and continuum shapes \citep{tomsick06,tomsick09,tomsicktbp12,rodriguez08,rodriguez09,rodriguez10}.

  From such spectroscopic and imaging efforts, it was possible to identify the nature of a large fraction of these IGR sources 
  \citep[and references therein]{chaty08,butler09,masetti07,masetti08,masetti12}. Among these sources $>$250 are AGN. 
  The Galactic sources include High-Mass X-ray Binaries, Low-Mass X-ray Binaries and Cataclysmic Variables (CVs), isolated neutron star 
  (NS) systems as well as supernova remnants and pulsar wind nebulae. Yet, the nature of one third of IGR sources has not been 
  identified. The identifications include a low number of unexpected hard X-ray sources like RS Canes Venatici (RS CVn), 
  and symbiotic binary systems \citep{bird10,masetti12,rodriguez10}, and thus searches for more such objects are warranted.

  So far, there have been four IGR sources tentatively identified as RS CVn binaries \citep{rodriguez10,masetti12,krivonos10}, but no 
  firm association with RS CVn systems and IGR sources exists. In general, RS CVn systems are the members of chromospherically active stars which include flare stars (dMe stars or UV Ceti 
  stars) and T Tauri stars as well. These systems are detached binaries having emission lines of CaII H \& K and H$\alpha$ lines in their 
  optical spectra. 
  The components of an RS CVn system can be G--M type stars within the luminosity classes of II--V \citep{kogleu07}. 
  This type of system is known to have strong X-ray and radio emission \citep{barbier05,osten07}. 
  They show solar like magnetic activities and flares \citep{hais90}.
  
  Symbiotic systems are long period interacting binaries consisting of a red giant star (G--M) and a hot companion surrounded by an 
  ionized nebula \citep{fried82,kenyon86,kenyon90,kenweb84}. The hot component is most often a white dwarf (WD) or a disk-accreting main 
  sequence star. On the other hand there are also systems having a NS as the compact companion \citep{miko97}. 
  In general, the optical spectra of symbiotic stars have photospheric absorption features and molecular bands of the cool giant and 
  emission lines of the HI Balmer series, HeI, HeII, [OIII] and [NeIII] which are produced in the surrounding nebula of the hot 
  component by its intense UV radiation \citep{kenfer87,cies94}. In addition, these systems can show hard, soft and supersoft X-ray emissions \citep{kennea09,muerset97}. 
  
  So far, eleven IGR sources have been suggested to be symbiotic systems according to their optical/infrared features and X-ray properties. 
  Among those identified, the symbiotic nature of IGR J10109$-$5746 (CD$-$57 3057; \citealp{masetti06,kennea09}), 
  IGR J12349$-$6434 (RT Cru; \citealt{masetti05,lunasoko07,kennea09}) and IGR J16194$-$2810 \citep{masetti07,ratti10} are well determined through 
  multiwavelength analyses. A nearby system (1.56 kpc, parallax measurement), IGR J15293$-$5609, has been classified as a 
  symbiotic binary based on the well determined surface temperature and radius of the companion star \citep{tomsicktbp12}. 
  On the other hand, the identifications of IGR J16358$-$4726 \citep{nespoli10},  and IGR J17497$-$2821 \citep{paizis07,paizis09} as 
  symbiotic systems are questionable. \cite{chaty08} suggest that IGR J16358$-$4726 could be a high mass X-ray binary. Likewise the classification 
  of IGR J17497$-$2821 is also controversial due to the suggestion of a black hole primary \citep{walter07,paizis07} would make the system very unusual among the 
  known symbiotics. IGR J16393$-$4643 was also claimed to be a symbiotic system based on the $K$-band spectrum of the candidate counterpart 
  2MASS J16390535$-$4642137 \citep{nespoli10} which was previously classified as a BIV--V type star using its optical/IR spectral energy 
  distribution \citep{chaty08}. Later, it was revealed that the refined X-ray position of IGR J16393$-$4643 was incompatible with 
  this controversial 2MASS source \citep{bodaghee12}.
  Candidate counterparts to IGR J11098$-$6457, IGR J17197$-$3010 and 1RXS J174607.8$-$213333 
  \citep{masetti09,masetti12,masetti08} are also tentatively claimed to be late M type giants based on comparison of their optical 
  spectra with the known symbiotic systems, but it was later shown that  the tentative optical counterpart identifications of 
  IGR J11098$-$6457 and IGR J17197$-$3010 are wrong \citep{tomsick09, luna12}. Finally,  IGR J16293$-$4603 was suggested to be a 
  candidate symbiotic system by \cite{ratti10} based solely on its optical/NIR photometry.
     
  In this work, we investigated the optical candidate counterparts of four IGR sources in the 4$^{th}$ Bird Catalog 
  \citep{bird10} with unknown X-ray classification: IGR J04069$+$5042, IGR J06552$-$1146, IGR J21188$+$4901 and IGR J22014$+$6034. 
  The fields of these sources were observed with {\emph{Chandra}}, and possible optical/infrared counterparts were 
  identified \citep[][hereafter T12]{tomsicktbp12}. All four sources have low Galactic latitudes (see Table~\ref{catalog}), 
  and high extinction limits the possible counterparts mostly to relatively nearby stellar objects. We performed optical spectroscopy 
  of bright candidates in the USNO and 2MASS catalogs, and also obtained images that were located near the brightest {\emph{Chandra}} 
  sources. This work is follow up of T12 that utilized \emph{Chandra} to find X-ray counterparts of IGR sources for further investigations.

  The paper is organized as follows. In Section 2, we summarize the observations and data reduction in the optical, in Section 3 we identify the types of 
  optical counterpart candidates of the IGR sources, and in the final section, we discuss the possible counterparts to each IGR source, and compare the M type candidates we found with earlier tentative symbiotic identifications.

\section{Observations and Data Reduction}

  The {\emph{INTEGRAL}} coordinates and the fluxes of the four sources we investigated are given in Table~\ref{catalog}. The 
  {\emph{Chandra}} localizations (See Table~\ref{chandra}) of the candidate counterparts allowed us to select sources to be observed 
  at the T\"{U}B\.{I}TAK Turkish National Observatory\footnote{\url{http://www.tug.tubitak.gov.tr}} (TUG) and at the MDM 
  Observatory. A detailed list of all detected possible {\emph{Chandra}} counterparts and their soft X-ray properties are reported 
  in T12. All of the X-ray sources detected by {\emph{Chandra}} in the field of each IGR source are shown in Figure~\ref{DSS}. 
  We also searched the online catalogs of 2MASS, USNO-B1.0 and USNO-A2.0 to find the IR and optical counterparts associated with the 
  {\emph{Chandra}} positions (See Table~\ref{candidates}).
    
  The spectroscopic observations of candidate counterparts for IGR J04069$+$5042 and IGR J06552$-$1146 were done at the MDM Observatory, 
  while IGR J21188$+$4901 and IGR J22014$+$6034 were observed at TUG.  
   
  \subsection{TUG Observations}
  The medium-resolution spectra of the candidate counterparts to the sources \I21188+4901 and \I22014$+$6034 were 
  obtained with the T\"{U}B\.{I}TAK Faint Object Spectrometer and Camera (TFOSC) which is mounted on the Russian-Turkish 
  1.5 m Telescope (RTT150) located at TUG, Antalya, Turkey. The camera is equipped with a 2048 $\times$ 2048, 
  15 $\mu$m pixel Fairchild 447BI CCD. It has a FOV of 13$\arcmin$ $\times$ 13$\arcmin$ with a pixel scale of 
  0\farcs 39 pixel$^{-1}$. Grism \#15, having an average dispersion of 3 \AA\ pixel$^{-1}$, and slit 67 
  $\mu$m (1$\farcs$24) were used providing a 3300--9000 \AA\ wavelength band.  
    
  All spectroscopic data, acquired on 2011, August 27 under good weather conditions, were reduced using the Long-Slit package of 
  MIDAS\footnote{\url{http://www.eso.org/projects/esomidas/}} to obtain 1D spectra of target sources. The spectra were 
  corrected for bias, flat-fielded and cleaned from the cosmic-ray hits with standard MIDAS routines. Wavelength 
  calibration of the spectra were done by Neon lamp spectrum. The heliocentric correction was applied to each spectrum after 
  the extraction was done. The spectrophotometric standard star BD $+$33$\degr$2642 \citep{oke90} was observed during 
  the observing night to get the flux-calibrated spectra of the candidate counterparts. 
  
  The equivalent widths (EW) of prominent absorption and emission lines are measured by the ALICE subroutine of MIDAS.

     
%
\subsection{MDM Observations}

  We used the OSMOS (Ohio State Multi-Object Spectrograph) with the MDM4K CCD on the 2.4 m Hiltner Telescope of the MDM Observatory 
  on 2011 January 11 and 13. The Volume Phase Holographic (VPH) transmission grating we utilized has a 1$\farcs$2 wide slit with a 
  wavelength coverage of 3900-6800 \AA\ at 3.5 \AA\ resolution. Conditions were clear during the observing runs. 
  Spectra of two {\emph{Chandra}} localized X-ray sources in the {\emph{INTEGRAL}} error circle of \I04069$+$5042, and of one X-ray 
  source for \I06552$-$1146 were obtained. The spectral reduction was performed using standard 
  IRAF\footnote{\url{http://iraf.noao.edu/}} routines. The flux calibration was done by \cite{oke83} standard stars. 
  We also obtained images in the $R$-band using the same CCD for sources that are not in the standard catalogs.

\subsection{\emph{Chandra} Observations}

  We used the {\emph{Chandra}} observations not only for determining optical counterpart candidate positions, but also obtaining soft
  X-ray properties which could aid in the identification process. We fitted the \emph{Chandra} 0.3--10 keV spectra with an absorbed 
  power-law model to determine the absorbed 0.3--2 keV and 2--10 keV flux for each source. We defined the hardness ratio as 
  $(F_{2-10 {\rm keV}} - F_{0.3-2 {\rm keV}}) / (F_{2-10 {\rm keV}} + F_{0.3-2 {\rm keV}})$. For details of spectral fitting, 
  background subtraction and flux calculation, see T12. For soft X-ray properties relevant for this work, see Table~\ref{chandra}.
\section{Results}
\label{results}   
\subsection{IGR J04069$+$5042}
   We performed optical spectroscopy of two possible candidates localized by {\emph{Chandra}} within the {\emph{INTEGRAL}} 
   error circle of 4\farcm2. 
   \\ \\
   2MASS J04064392$+$5044469 (\#1): The optical spectrum includes prominent absorptions of NaI D resonance lines, MgI b and CaI triplet 
   lines (See Fig.~\ref{04069s}). Although distinguishing features in the classification region are mostly at the noise level, 
   the spectrum is very similar to the K type main sequence star HD 283916 (SAO 76803) in \citet{jacoby84} and to the sources 
   within the same class in \citet{roos96}. Indeed the presence of MgH bands at around $\lambda$4780 and $\lambda$5200, 
   seen only in the spectra of main sequence stars, supports the luminosity classification. On the other hand a weak but noticable 
   H$\alpha$ emission line (EW = 1.07 \AA, see Table~\ref{MDMwidths}) which could be due to the chromospheric activities, stellar 
   winds and/or interactions  with the environment of the secondary (or primary) indicates an active star. Therefore we classify this 
  source as a K5-7 Ve type star based on its spectral lines and EW measurements.
   \\ \\
   2MASS J04064872$+$5039316 (\#2): The spectrum is dominated by neutral Hydrogen lines (H$\delta$, H$\gamma$, H$\beta$ and H$\alpha$) 
   and several metallic lines  of FeI, SrII, MgI and MgII (See Fig.~\ref{5039316}). The CaI triplet lines that are used as luminosity 
   indicators are very weak (See Table.~\ref{MDMwidths}) as seen in the spectra of F type stars \citep{montes98,cayrel96}. In addition 
   the G band of CH at $\lambda$4300 that becomes visible for the spectral types later than F4 and increases in strength with 
   decreasing temperature \citep{gray09} is one of strongest features in the blue region (4000--5000 \AA) and allows us to exclude 
   the earlier spectral classes. Comparing the line width ratio of $\lambda$4077 (SrII) to $\lambda$4226 (CaI) that is a sensitive 
   criteria for luminosity \citep{morgan43}, and the EW measurements of Balmer lines (See Table~\ref{MDMwidths}) to the values in 
   \citet{jasc95} and in \citet{eaton95} we suggest that this source could be an F5--8 type star within the luminosity class of 
   III--V.  
\subsection{\I06552$-$1146} 
   Optical spectroscopy was only obtained for candidate \#1 (see Fig.~\ref{DSS}) which is $\sim$0\farcs8 away from its 
   {\emph{Chandra}} counterpart. Candidate \#2 is 2\farcs8 away from the nearest source in 2MASS and USNO catalogs. As it can 
   also be seen in the $R$-band image of this source (Fig.~\ref{MDM06}), the {\emph{Chandra}} source cannot be clearly identified 
   with an optical counterpart. Likewise, candidate \#3 was not listed in Table~\ref{catalog} since its coordinates are not associated 
   with any optical or IR counterparts in the catalogs. However, a weak source in the {\emph{Chandra}} error circle is visible in the 
   $R$-band image of the field (Fig.~\ref{MDM06}). This source has an R magnitude of 22.17 $\pm$ 0.14.
   \\ \\
   2MASS J06545833$-$1149119 (\#1): The flux-calibrated spectrum of candidate \#1 shows FeI, MgI and CaI absorption lines 
   (See Fig.~\ref{06552s}) and a weak (EW = 0.45 \AA) H$\alpha$ emission line (see Table~\ref{MDMwidths}). The characteristics 
   of the spectrum are very similar to those of candidate \#1 to IGR J04069$+$5042 in terms of the existence of the spectral lines 
   and the molecular MgH band. To give a spectral range, we searched the spectral atlases and libraries having a similar spectrum to 
   that of 2MASS J06545833$-$1149119. By comparing the spectra of the sources listed in \citet{jacoby84}, \citet{roos96} and in 
   \citet{gray09} to candidate \#1 we suggest that it would be a main sequence K5--8 type emission-line star. 
\subsection{\I21188$+$4901}
   This source is tentatively classified as a transient in \citet{bird10} due to its highly 
   variable X-ray flux in 20--40 keV energy band. Two candidates which are quite faint in optical band have been 
   observed (Fig.~\ref{DSS}) to search for the optical counterpart to \I21188$+$4901.
   \\ \\
   2MASS J21182288$+$4906259 (\#1): The optical spectrum of this source contains both broad molecular TiO bands with a red 
   continuum and an H$\alpha$ emission line with EW = 1.9 \AA. Although EW values of H$\gamma$ and H$\beta$ lines could not be 
   determined due to their low signal to noise, they are apparent in emission. These spectral features are indicative of a late-type 
   star spectrum with chromospheric activity. CaI and CaII triplet lines, used as luminosity indicator for M type stars \citep{jasc95}, 
   MgI lines and telluric absorption bands are the other identified features in the spectrum (See Fig.~\ref{21188n1}). We point out 
   that the spectral apperance of this candidate is very similar to that of the counterparts of 1RXS J174607.8$-$213333 and 
   IGR J11098$-$6457 which are identified as an M2--4 \citep{masetti08} and an M2-type \citep{masetti09} symbiotic giants respectively. 
   However this source does not have the clear spectral features of a red giant star. In addition 2MASS magnitudes are too faint to be 
   a nearby giant. Therefore we searched for the late-type stars spectra in the atlases of \citet{gunn83}, \citet{jacoby84} and 
   \citet{silvacor92} to compare with our flux-calibrated spectrum and we determine that the spectral class of this candidate could 
   be a M3--5 type active main sequence star. 
   \\ \\
   2MASS J21183906$+$4858049 (\#2): This candidate is the only source having a {\emph{Chandra}} counterpart within the 
   {\emph{INTEGRAL}} error circle. The spectrum shown in Fig.~\ref{21188n2} has typical features of a late-K/mid-M type star 
   \citep{gray09} without any emission lines. In addition the molecular CaOH band that becomes visible for dwarfs having spectral 
   class later than K7 and MgH bands are prominent features in the spectrum. The weakness and/or absence of the Balmer lines 
   (See Table~\ref{TUGwidths}) support the late-type spectral classification. Therefore, considering characteristics of the spectrum 
   we suggest a K7--M4 type main sequence classification for this source. 
\subsection{\I22014$+$6034}
   We obtained optical spectroscopy of four stars that are detected with {\emph{Chandra}} in 2MASS and USNO catalogs.
   Two of these candidate optical/IR counterparts lie within the $5^{\prime}\!.4$-radius {\emph{INTEGRAL}} error
   circle as seen in Figure~\ref{DSS}.
   \\ \\
   2MASS J22002116$+$6033420 (\#1): The flux-calibrated spectrum is very similar to candidate \#2 of IGR J21188$+$4901 that is 
   classified as a K7--M4 type main sequence star. It is dominated by the molecular MgH, CaOH and CH-G bands in addition to MgI b 
   triplet lines (See Fig.~\ref{22014n1}). The emission lines of H$\alpha$ and CaII H \& K at $\lambda$3968, $\lambda$3934 
   (See Table~\ref{TUGwidths}), indicators of chromospheric activity, are also detected. 
   \\ \\  
   2MASS J22010934$+$6034088 (\#2): This IR source is the optically brightest candidate to the X-ray position of \I22014$+$6034.
   However, the optical spectrum contains no significant features related to an X-ray source. The most prominent absorptions in the blue 
   region of the spectrum are of CaII H \& K and Balmer series lines. Metallic lines of FeI and MgI b triplet are also visible.
   The flux-calibrated spectrum of  2MASS J22010934$+$6034088 is similar to that of A--G type stars due to its decaying blue 
   continuum towards the longer wavelengths (see Fig.~\ref{22014_n2}). On the other hand the presence of CH-G band, and the weakness of 
   Balmer lines (See Table~\ref{TUGwidths}) while compared to A-type stars indicate that this source must be in F--G spectral range. 
   By comparing EW values of CaII triplet lines in near IR region to those of late-type stars \citep{jasc95,zhou91}, we classify this 
   candidate as an F5-G5 type III-V star. 
   \\ \\
   2MASS J22015416$+$6038094 (\#3): This bright IR star is located at edge of the IGR error circle. It does not have an 
   optical counterpart in USNO-A2.0 catalog. Although the flux-calibrated spectrum contains nearly most of the features detected in 
   2MASS J22010934$+$6034088, they are different regarding the strength of the lines (See Table~\ref{TUGwidths}). The identification 
   spectrum of the candidate \#3 is shown in Figure~\ref{22014_n3}. The Balmer lines of H$\delta$ and H$\gamma$ are extremely weak or 
   absent, compared to H$\alpha$ line (EW = $\sim$2.3 \AA) and to CH-G band. The blue region of the spectrum which is dominated by metallic 
   lines of MgI, MgII and FeI displays typical spectral features of a late-type star. We derive a spectral class of G III--IV based 
   upon the prominent features for this source.
   \\ \\
   2MASS J22020837$+$6030425 (\#4): The coordinates of the source are coincident with the positions of USNO-B1.0 1505$-$0322924 and 
   USNO-A.2 1500$-$08693376 although they fall just outside the {\emph{INTEGRAL}} error circle (6\farcm06). The flux-calibrated 
   4000--9000 \AA\ spectrum shows typical features of a M type star (see Fig.~\ref{22014_n4}) dominated by strong molecular bands 
   of TiO \citep{gray09}. The emissions of Balmer series lines (H$\gamma$, H$\beta$, H$\alpha$) superposed with TiO bands and the 
   presence of CaI, CaII, MgI b lines are in favor of the late-type spectral classification (See Table~\ref{TUGwidths}). Since 
   determining exact spectral class of late-type stars is rather complicated due to the contaminated blue part of the 
   spectrum, we used spectrophotometric atlases of \citet{gunn83}, \citet{jacoby84} and the library of \citet{silvacor92} to secure 
   the spectral and luminosity classification. Comparing our spectrum with the late-type stars in the atlases and libraries we 
   classify 2MASS J22020837$+$6030425 as a main sequence M2--5 type chromospherically active star.
\subsection{X-ray results}

  The four IGR sources we investigated in this work are reported as sources for which none of the {\emph{Chandra}} sources in the 
  field of view stands out as a candidate counterpart from a likelihood analysis just based on X-rays as described in T12. This 
  likelihood analysis calculates the spurious association probability as:
\begin{equation}
P = 1-e^{-N(>F_{2-10 {\rm keV}}) \, \pi \,  \theta^{2}_{search}}
\label{prob}
\end{equation}
  where $N$($>$$F$$_{2-10 {\rm keV}}$) = 9.2 ($F_{2-10 {\rm keV}}/10^{-13})^{-0.79}$ deg$^{-2}$ \citep[][T12]{Sugizaki01}, and $F_{2-10 {\rm keV}}$ is the absorbed 2-10 keV flux  
  in ergs cm$^{-2}$ s$^{-1}$. $\theta$$_{search}$ is the radius of the search region in units 
  of degrees and equals to 90$\%$ confidence for the sources inside the {\emph{INTEGRAL}} error circle (See Table~\ref{catalog}). 
  For sources outside the error circle, it indicates the angular separation from the center of the {\emph{INTEGRAL}} error circle. 
  The spurious association probabilities of all sources are given in Table~\ref{chandra}.

  For most of the candidates the spurious association probability is very high. The notable exceptions are candidate \#3 of 
  \I06552$-$1146, candidate \#2 of \I21188$+$4901, and candidate \#3 of \I22014$+$6034. However the likelihood analysis does not take 
  into account the variability. The soft X-ray measurements were obtained at an arbitrary epoch, and {\emph{INTEGRAL}} analysis shows that 
  the maximum hard X-ray flux and the average hard X-ray flux are different for these sources indicating variability 
  (See Table~\ref{catalog}, T12, \citealp{bird10}). In fact, IGR J04069$+$5042 and IGR J06552$-$1146 have bursticity\footnote{Bursticity can be defined as the ratio of the maximum significance on any timescale, compared to the average significance \citep{bird10}.}
  impact grater than 1.1 and for IGR J21888$+$4901 it is greater than 4, indicating a highly variable nature of the source in X-rays 
  \citep{bird10}. Therefore likelihood analysis based on X-ray data alone may fail in revealing the true counterpart if the X-ray 
  observations take place when the source is weak. It is still possible that one of the detected {\emph{Chandra}} sources is the 
  correct counterpart.

\section{Discussion}

  Below, we discuss the candidate counterparts for each IGR source in terms of their optical spectra, 
  {\emph{Chandra}} soft X-ray properties and relative probabilities to find out the candidates that are more 
  likely to be the IGR source. 

\subsection{IGR J04069$+$5042} 

  According to the Equation~\ref{prob}, the spurious association probabilities of candidates \#1 and \#2 are $>$68\%, 
  and $>$ 99.99\% respectively, and neither of them was detected in the 2--10 keV band with {\emph{Chandra}}. Among these two sources, 
  candidate \#2 has typical features of a late F-type star. On the other hand, candidate \#1 shows H$_\alpha$ emission, 
  which makes it more likely to be the correct counterpart. 

\subsection{IGR J06552$-$1146}

  Candidate \#1 is 4\farcm14 away from the center of the {\emph{INTEGRAL}} error circle and the probability of spurious association 
  is 88\%. The spectrum shows evidence of an H$_\alpha$ emission line and it is similar to that of candidate \#1 for \I04069$+$5042. 
  The other sources with high ACIS count rates are too dim in the optical to get a spectrum. Their spurious association probabilities 
  are 33--98\% and 14--29\% for candidates \#2 and \#3 respectively. Candidate \#3 in fact has positive hardness ratio, and the relative 
  probability is quite close to the cut-off for a {\emph{Chandra}}/{\emph{INTEGRAL} association in T12. 

\subsection{IGR J21888$+$4901}

  Candidate \#1 is outside the 90\% confidence error circle of {\emph{INTEGRAL}} with spurious association probability of $>$99.96\%. 
  It has been detected with {\emph{Chandra}} in the 2--10 keV band, and is an active M type star. Candidate \#2 is the only soft X-ray 
  source (detected in 0.3--2 keV energy band) in the {\emph{INTEGRAL}} error circle with a probability of $>$15\% but it is not detected with {\emph{Chandra}} in the 
  2--10 keV band. Candidate \#2 shows properties of a late-type (K7--M4) main sequence star. Based on probabilities, candidate \#2 is a 
  more likely counterpart than candidate \#1.

\subsection{IGR J22014$+$6034}
  The first candidate is the brightest \emph{Chandra} source in the vicinity of the IGR source, however it is out of the 90\% 
  confidence error circle of {\emph{INTEGRAL}}. The spectrum is typical of an active late-type star. The probability of spurious 
  association is $>$95\% and it is detected in the 2--10 keV band with {\emph{Chandra}}, however the source is very soft.

  The optical spectra of candidates \#2 and \#3 show that they are late-type main sequence stars without emission lines. 
  The probabilities of spurious associations for these sources are $>$88\% and $>$36\% respectively. 

  The optical spectrum of candidate \#4 shows features of an active M type main sequence star. It is located just outside the 
  error circle of {\emph{INTEGRAL}}. The probability is $>$99.98\%, and it is not detected with {\emph{Chandra}} in the 
  2--10 keV band. 


\subsection{IGR sources identified as symbiotic systems}
\label{subsec:symbio}

  At an early phase of our analysis, we compared our optical spectra with those published in \citet{masetti08, masetti09, masetti12} 
  and realized the similarities of candidate \#1 of IGR J21888$+$4901 and candidate \#4 of IGR J22014$+$6034, the M type stars with 
  broad TiO bands, to those identified as symbiotic stars. However, these sources do not show HeI, HeII and [OIII] in emission
  \citep{kenyon86}, and their infrared colors are not consistent with the colors of symbiotics in the $J-H$ vs $H-Ks$ diagram as given 
  by \citet{phil07} and \citet{Corradi08}. Our candidates are more probably active main sequence stars e.g., dMe stars \citep{pettersenhawley89} 
  rather than being giants. To illustrate this we plotted the NIR color-color diagram of M type TUG candidates (filled red
  circles) along with colors of several other sources in Figure~\ref{symbio}. The plot includes validated IGR symbiotics 
  (IGR J10109$-$5746, IGR J12349$-$6434 and IGR J16194$-$2810 with filled blue circles) and the all known symbiotics 
  (small, empty, purple circles), the boundaries that limit stellar (S) and dusty (D) type symbiotics \citep[see][for details of S 
  and D type symbiotics]{pereira09} and the loci of main sequence stars and red giant stars. The colors of IGR sources, 
  tentatively classified as symbiotic systems but later shown to have the incorrect counterpart candidates, are also placed in 
  Figure~\ref{symbio}. The initial classifications of the counterparts to IGR J11098$-$6457, IGR J17197$-$3010 (empty red circles) 
  and IGR J16393$-$4643 (empty orange circle) were done with optical and/or IR spectral analyses. Since the refined X-ray positions of 
  these IGR sources were not related to the suggested counterparts, the symbiotic identifications were ruled out. Note that the colors 
  of incorrect counterparts can easily be associated with the main sequence stars like the TUG candidates.
  
  Two other candidates were also tentatively claimed to be symbiotic binaries, IGR J16358$-$4726, IGR J17497$-$2821, but subsequent 
  analyses made the symbiotic interpretation questionable (see Introduction for details). These sources are shown as "IGR Sym, 
  questionable" in Fig.~\ref{symbio} (filled brown circles), and have colors consistent with dusty symbiotic systems. The colors of the 
  infrared counterpart of IGR J16293$-$4603 (filled orange circle) are also consistent with being a symbiotic system as expected, 
  since the initial identification was done using $g$, $r$, and $i$ colors by \cite{ratti10}. We also show IGR J15293$-$5609, 
  classified as a symbiotic with a $K$ type companion (blue star). This source falls slightly out of the region of validated symbiotics. 
  On the other hand, symbiotics with $K$ type companions are rare, and occupy lower $J-H$ values \citep{pereira09} just like 
  IGR J15293$-$5609. These are called yellow symbiotic systems which are indicated separately in
  Figure~\ref{symbio} (filled, dark yellow circles). We also note that reddening can move the sources along the diagonal lines 
  shown in the figure, and the large uncertainty in $A_{V}$ of IGR J15293$-$5609 (T12) could easily move this source to the confirmed 
  region. Finally we show the colors of all other candidates that we identify as main sequence stars (with triangles), 
  and as expected most of them lie along the locus of main sequence stars.

  This analysis shows that near infrared color diagram provides a quick method to double check identifications of symbiotics with 
  some caveats. First of all, many symbiotic systems have normal NIR colors (see confirmed symbiotics, and the locus of red giant 
  branch stars in Fig.~\ref{symbio}). NIR colors may be enough to claim a giant identification, but it does not mean that the system 
  is a symbiotic. Second, most Galactic IGR sources are in the plane with high extinction (including the sources we analyzed, all with 
  $A_{V} > 3$) which could move main sequence stars into the range of symbiotics. We could not include colors of 1RXS J174607.8$-$213333 
  \citep{masetti08} on this plot because it was not in the near infrared catalogs, but that identification also relied only on low 
  resolution optical spectrum which requires further analysis. As mentioned in \citet{masetti08}, it could also be a spurious detection.

\section{Conclusions}

  We tabulate the the properties of all of the investigated sources in Table~\ref{summary}. We found several chromospherically active 
  late-type stars that might be the counterparts of the IGR sources. A special type of active stars, the RS CVn systems, significantly contribute to the population of X-ray sources above 2 keV \citep{motch10}. After the tentative 
  identification of IGR J08023$-$6954 as an RS CVn system by \citet{rodriguez10}, two more IGR sources were identified in this class 
  of binaries with active coronae \citep{masetti12}. These sources are variable in hard X-rays \citep{barbier05,osten07}, 
  and therefore can be detected with \emph{INTEGRAL}. Given the similarity of the optical spectra, candidate \#1 of 
  IGR J04069$+$5042 or candidate \#1 of IGR J06552$-$1146 may be RS CVn if either of them is the actual counterpart.

  The total number of IGR systems claimed to be symbiotics is eight. Including IGR J15293$-$5746, the number of multiple method confirmed symbiotics 
  is four. For our candidates with deep TiO bands, we cannot confirm the symbiotic nature as their near infrared colors are not consistent 
  with known symbiotics. Given that the active dMe stars are ubiquiotus in our Galaxy, some of the suggested counterparts to the 
  IGR sources claimed as symbiotics may as well be active main sequence stars. We stress that low resolution optical spectra may not 
  always provide right identification for these type of systems, and further high resolution, multiwavelength and timing 
  observations may be required to understand the true nature of these sources. 

\begin{acknowledgements}
  M.\"O.A. acknowledges support from T\"{U}B\.{I}TAK, The Scientific and Technological Research Council of Turkey, through the research 
  project 109T736. J.A.T. acknowledges partial support from NASA through {\em Chandra} Award Number GO1-12046X issued by the 
  {\em Chandra} X-ray Observatory Center, which is operated by the Smithsonian Astrophysical Observatory under NASA contract NAS8-03060. We thank 
  the Turkish National Observatory of T\"{U}B\.{I}TAK for running the optical facilities. We thank Dr. Timur \c{S}ahin for fruitful 
  discussions.  
\end{acknowledgements}

\clearpage
\begin{deluxetable*}{ccccccccc}
\tablewidth{0pt}
\tablecaption{The X-ray measurements of IGR sources given in \citet{bird10}\label{catalog}}
\tablehead{
\colhead{IGR Name} & \colhead{RA(J2000)} & \colhead{DEC(J2000)} & \colhead{\textit{l}\tablenotemark{a}}&\colhead{\textit{b}\tablenotemark{b}} &\colhead{Error Radius\tablenotemark{c}} & 
\colhead{Flux$_{20-40 \ keV}$} & \colhead{Flux$_{40-100 \ keV}$} & \colhead{Peak Flux\tablenotemark{d}$_{20-40 \ keV}$}}
\startdata
J04069$+$5042 & 04$^{h}$06$^{m}$55$\fs$0  & $+$50$\degr$42$\farcm$1 & 151.43&$-$1.03&4$\farcm$2 & 1.6$\pm$0.3 & \textless1.0 & 2.0$\pm$0.3\\
J06552$-$1146 & 06$^{h}$55$^{m}$10$\fs$0  & $-$11$\degr$46$\farcm$2 & 223.85&$-$4.52&4$\farcm$3 & 1.1$\pm$0.3 & \textless1.0 & 2.6$\pm$1.0\\
J21188$+$4901 & 21$^{h}$18$^{m}$48$\fs$0  & $+$49$\degr$01$\farcm$0&91.27 &$-$0.33&4$\farcm$3 & \textless0.2 & \textless0.4 & 4.5$\pm$1.3\\
J22014$+$6034 & 22$^{h}$01$^{m}$27$\fs$0  & $+$60$\degr$34$\farcm$0&103.49 &$+$4.28&5$\farcm$4 & \textless0.2\tablenotemark{e} &\textless0.4 &\nodata\\
\enddata
\tablecomments{The fluxes are given in units of mCrab. A flux of 1 mCrab in the 20--40 keV energy range corresponds to 
7.57$\times$10$^{-12}$ erg cm$^{-2}$ s$^{-1}$ while it is 9.42$\times$10$^{-12}$ erg cm$^{-2}$ s$^{-1}$ for 40--100 keV band.}
\tablenotetext{a}{The Galactic Longitude of the source in degrees.}
\tablenotetext{b}{The Galactic Latitude of the source in degrees.}
\tablenotetext{c}{The radius of the 90\% confidence IGR circle.}
\tablenotetext{d}{The peak flux of the sources which show variability in the 20--40 keV energy band. For \I22014$+$6034 no variation 
detected in this range.}
\tablenotetext{e}{The source is detected in 17--30 keV band, \citet{bird10}.}
\end{deluxetable*}
\clearpage
%
\begin{deluxetable*}{cccccl}
\tablecolumns{6}
\tablewidth{0pc}
\tablecaption{\emph{Chandra} Localizations, Counts, Hardness Ratio, and The Probability of Spurious Associations\label{chandra}}
\tablehead{
\multicolumn{6}{c}{\textbf{IGR J04069$+$5042}}\\ 
\colhead{$^{Candidate \#}$Chandra Name\tablenotemark{a}} & \colhead{RA(J2000)} & \colhead{DEC(J2000)} & 
\colhead{ACIS Counts} & \colhead{Hardness\tablenotemark{b}} & \colhead{$P$ (\%)\tablenotemark{c}}}\\
\startdata
$^{1}$CXOU J040643.9$+$504446 & 04$^{h}$06$^{m}$43$\fs$99 & $+$50$\degr$44$\arcmin$46$\farcs$2  & 5.4  &$-$1.19$\pm$1.10 & $>$68\\
$^{2}$CXOU J040648.7$+$503931 & 04$^{h}$06$^{m}$48$\fs$69 & $+$50$\degr$39$\arcmin$31$\farcs$2  & 3.4 &  $-$1.30$\pm$1.65 & $>$99.99 \\
\cutinhead{\textbf{IGR J06552$-$1146}}
$^{1}$CXOU J065458.3$-$114911 & 06$^{h}$54$^{m}$58$\fs$31 & $-$11$\degr$49$\arcmin$11$\farcs$3  & 8.5 & $-$0.75$\pm$0.70 & $>$83 \\
$^{2}$CXOU J065523.6$-$114601 & 06$^{h}$55$^{m}$23$\fs$65 & $-$11$\degr$46$\arcmin$01$\farcs$1  & 13.4 & $-$0.62$\pm$0.46 & 33-98 \\
$^{3}$CXOU J065529.5$-$114900 & 06$^{h}$55$^{m}$29$\fs$56 & $-$11$\degr$49$\arcmin$00$\farcs$1  & 35.5 & $+$0.02$\pm$0.22 & 14-29\\
\cutinhead{\textbf{IGR J21188$+$4901}}
$^{1}$CXOU J211822.9$+$490627 & 21$^{h}$18$^{m}$22$\fs$86 & $+$49$\degr$06$\arcmin$26$\farcs$7 & 7.1  & $-$0.78$\pm$0.82 & $>$99.96 \\
$^{2}$CXOU J211839.1$+$485806 & 21$^{h}$18$^{m}$39$\fs$09 & $+$48$\degr$58$\arcmin$05$\farcs$5 & 4.3 &  $+$0.61$\pm$1.00 & $>$15\\
\cutinhead{\textbf{IGR J22014$+$6034}}
$^{1}$CXOU J220021.1$+$603342 & 22$^{h}$00$^{m}$21$\fs$14 & $+$60$\degr$33$\arcmin$41$\farcs$8 & 45.8 & $-$40.91$\pm$0.25 & $>$95\\         
$^{2}$CXOU J220109.3$+$603409 & 22$^{h}$01$^{m}$09$\fs$34 & $+$60$\degr$34$\arcmin$08$\farcs$7  & 14.4& $-$0.79$\pm$0.46 & $>$88\\          
$^{3}$CXOU J220154.1$+$603809 & 22$^{h}$01$^{m}$54$\fs$13 & $+$60$\degr$38$\arcmin$09$\farcs$5 & 31.7 & $-$0.61$\pm$0.27 & $>$36\\
$^{4}$CXOU J220208.4$+$603042 & 22$^{h}$02$^{m}$08$\fs$35 & +60$\degr$30$\arcmin$42$\farcs$4 & 10.7 & $-$1.15$\pm$0.70 & $>$99.99\\
\enddata
\tablenotetext{a}{All {\emph{Chandra}} information is reported by T12}
\tablenotetext{b}{Hardness ratio is defined as $(C_{2-10 keV} - C_{0.3-2 keV}) / (C_{2-10 keV} + C_{0.3-2 keV})$ where $C$ is the count 
rate in the given band. After background subtraction in {\emph{Chandra}}, the hardness ratios may become negative (T12).}
\tablenotetext{c}{See Equation~\ref{prob} and T12.}
\end{deluxetable*}
\clearpage
\begin{deluxetable*}{lccccccc}
\tablecolumns{8}
\tablewidth{0pc}
\tablecaption{Observed Possible Optical/Ir Counterparts To The Sources\label{candidates}}
\tablehead{
\multicolumn{8}{c}{\textbf{IGR J04069$+$5042}}\\
\colhead{$^{Candidate \#}$Catalog/Source Name\tablenotemark{a}}& \colhead{RA(J2000)}& \colhead{DEC(J2000)} & 
\colhead{Distances\tablenotemark{b}} & \multicolumn{4}{c}{Magnitudes}}
\startdata
$^{1}$2MASS J04064392$+$5044469 & 04$^{h}$06$^{m}$43$\fs$92 & $+$50$\degr$44$\arcmin$46$\farcs$95  &3\farcm 20/1\farcs00&\multicolumn{4}{c}{J=14.40 \ \ H=13.68 \ \ K$_s$=13.45 \ }\\
\ USNO-A2.0 1500$-$08672098 & 04$^{h}$06$^{m}$43$\fs$94 & $+$50$\degr$44$\arcmin$47$\farcs$39 & 3\farcm 23/1\farcs28 &\multicolumn{4}{c}{B=18.60 \ \ R=16.60}\\
\ USNO-B1.0 1505$-$0322270  & 04$^{h}$06$^{m}$43$\fs$92 & $+$50$\degr$44$\arcmin$47$\farcs$32 & 3\farcm 21/1\farcs29 &\multicolumn{4}{c}{B$_1$=19.06 R$_1$=16.32 B$_2$=18.49 R$_2$=16.71 I=15.42}\\
\\
\tableline
\\
$^{2}$2MASS J04064872$+$5039316 & 04$^{h}$06$^{m}$48$\fs$73 & $+$50$\degr$39$\arcmin$31$\farcs$63  &2\farcm 76/0\farcs58&\multicolumn{4}{c}{J=11.44 \ \ H=11.13 \ \ K$_s$=11.06 \ }\\
\ USNO-A2.0 1350$-$04221566 & 04$^{h}$06$^{m}$48$\fs$74 & $+$50$\degr$39$\arcmin$31$\farcs$81 & 2\farcm 75/0\farcs78 &\multicolumn{4}{c}{B=13.20 \ \ R=12.30}\\
\ USNO-B1.0 1406$-$0110577  & 04$^{h}$06$^{m}$48$\fs$73 & $+$50$\degr$39$\arcmin$31$\farcs$76 & 2\farcm 76/0\farcs69 &\multicolumn{4}{c}{B$_1$=13.56 R$_1$=12.15 B$_2$=13.37 R$_2$=12.69 I=11.82}\\
\cutinhead{\textbf{IGR J06552$-$1146}}
\\
$^{1}$2MASS J06545833$-$1149119 & 06$^{h}$54$^{m}$58$\fs$34 & $-$11$\degr$49$\arcmin$12$\farcs$00  &4\farcm 14/0\farcs79&\multicolumn{4}{c}{J=14.04 \ \ H=13.83 \ \ K$_s$=13.32 \ }\\
\ USNO-A2.0 0750$-$03064876 & 06$^{h}$54$^{m}$58$\fs$35 & $-$11$\degr$49$\arcmin$11$\farcs$80  &4\farcm 14/0\farcs74&\multicolumn{4}{c}{B=16.60 \ \ R=15.40 }\\
\ USNO-B1.0 0781$-$0141882  & 06$^{h}$54$^{m}$58$\fs$33 & $-$11$\degr$49$\arcmin$11$\farcs$89 & 4\farcm 14/0\farcs64 &\multicolumn{4}{c}{B$_1$=16.95 R$_1$=15.43 B$_2$=17.48 R$_2$=15.34 I=14.60}\\
\\
\cutinhead{\textbf{IGR J21188$+$4901}}\\ 
$^{1}$2MASS J21182288$+$4906259 & 21$^{h}$18$^{m}$22$\fs$89 & $+$49$\degr$06$\arcmin$25$\farcs$96 & 6\farcm 81/0\farcs78 &\multicolumn{4}{c}{J=13.01 \ \ H=12.45 \ \ K$_s$=12.17 \ }\\
\ USNO-A2.0 1350$-$13821833 & 21$^{h}$18$^{m}$22$\fs$89 & $+$49$\degr$06$\arcmin$26$\farcs$39 & 6\farcm 82/0\farcs40 &\multicolumn{4}{c}{B=18.20 \ \ R=16.30 \ }\\
\ USNO-B1.0 1391$-$0394771  & 21$^{h}$18$^{m}$22$\fs$88 & $+$49$\degr$06$\arcmin$26$\farcs$31 & 6\farcm 82/0\farcs44 &\multicolumn{4}{c}{B$_1$=18.33 R$_1$=16.01 B$_2$=18.24 R$_2$=16.06 I=14.51}\\
\\
\tableline
\\
$^{2}$2MASS J21183906$+$4858049 & 21$^{h}$18$^{m}$39$\fs$07 & $+$48$\degr$58$\arcmin$04$\farcs$92 & 3\farcm 27/0\farcs61&\multicolumn{4}{c}{J=14.41 \ \ H=13.78 \ \ K$_s$=13.47 \ }\\
\ USNO-A2.0 1350$-$13829476 & 21$^{h}$18$^{m}$39$\fs$00 & $+$48$\degr$58$\arcmin$05$\farcs$32 & 3\farcm 26/0\farcs91&\multicolumn{4}{c}{B=19.00 \ \ R=17.00 \ }\\
\ USNO-B1.0 1389$-$0393707  & 21$^{h}$18$^{m}$39$\fs$06 & $+$48$\degr$58$\arcmin$05$\farcs$28 & 3\farcm 26/0\farcs37& \multicolumn{4}{c}{B$_1$=18.57 R$_1$=17.12 B$_2$=18.66 R$_2$=16.69 I=15.91}\\
\cutinhead{\textbf{IGR J22014$+$6034}} \\ 
$^{1}$2MASS J22002116$+$6033420 & 22$^{h}$00$^{m}$21$\fs$17 & $+$60$\degr$33$\arcmin$42$\farcs$06 & 8\farcm 09/0\farcs34 &\multicolumn{4}{c}{J=11.66 \ \ H=10.99\ \ K$_s$=10.82 \ }\\         
\ USNO-A2.0 1500$-$08655368 & 22$^{h}$00$^{m}$21$\fs$17 & $+$60$\degr$33$\arcmin$42$\farcs$11 & 8\farcm 09/0\farcs38 &\multicolumn{4}{c}{B=15.50 \ \ R=13.7 \ }\\
\ USNO-B1.0 1505$-$0321733  & 22$^{h}$00$^{m}$21$\fs$27 & $+$60$\degr$33$\arcmin$42$\farcs$44 & 8\farcm 08/1\farcs13 &\multicolumn{4}{c}{B$_1$=15.41 R$_1$=13.61 B$_2$=15.19 R$_2$=13.38 I=12.29}\\
\\
\tableline
\\
$^{2}$2MASS J22010934$+$6034088 & 22$^{h}$01$^{m}$09$\fs$34 & $+$60$\degr$34$\arcmin$08$\farcs$81  &2\farcm 17/0\farcs11&\multicolumn{4}{c}{J=10.62 \ \ H=10.39 \ \ K$_s$=10.32 \ }\\         
\ USNO-A2.0 1500$-$08672098 & 22$^{h}$01$^{m}$09$\fs$29 & $+$60$\degr$34$\arcmin$08$\farcs$12 & 2\farcm 18/0\farcs67 &\multicolumn{4}{c}{B=12.70 \ \ R=11.00}\\
\ USNO-B1.0 1505$-$0322270  & 22$^{h}$01$^{m}$09$\fs$28 & $+$60$\degr$34$\arcmin$08$\farcs$00 & 2\farcm 18/0\farcs82 &\multicolumn{4}{c}{B$_1$=12.57 R$_1$=11.00 B$_2$=11.72 R$_2$=10.86 I=10.69}\\          
\\
\tableline
\\
$^{3}$2MASS J22015416$+$6038094 & 22$^{h}$01$^{m}$54$\fs$16 & $+$60$\degr$38$\arcmin$09$\farcs$48 & 5\farcm 33/0\farcs23 &\multicolumn{4}{c}{J=11.50 \ \ H=11.11 \ \ K$_s$=10.94 \ }\\
\ USNO-B1.0 1506$-$0321404  & 22$^{h}$01$^{m}$53$\fs$99 & $+$60$\degr$38$\arcmin$09$\farcs$60 & 5\farcm 32/1\farcs02 &\multicolumn{4}{c}{B$_1$=14.59 R$_1$=11.77 B$_2$=-    R$_2$=12.46 I=11.74}\\
\\
\tableline
\\
$^{4}$2MASS J22020837$+$6030425 & 22$^{h}$02$^{m}$08$\fs$37 & $+$60$\degr$30$\arcmin$42$\farcs$50 & 6\farcm 06/0\farcs18 &\multicolumn{4}{c}{J=12.59 \ \ H=11.95 \ \ K$_s$=11.71 \ }\\
\ USNO-A2.0 1500$-$08693376 & 22$^{h}$02$^{m}$08$\fs$28 & $+$60$\degr$30$\arcmin$42$\farcs$83 & 6\farcm 05/0\farcs68 &\multicolumn{4}{c}{B=17.50 \ \ R=15.50 \ }\\
\ USNO-B1.0 1505$-$0322924  & 22$^{h}$02$^{m}$08$\fs$38 & $+$60$\degr$30$\arcmin$42$\farcs$77 & 6\farcm 06/0\farcs43 &\multicolumn{4}{c}{B$_1$=17.37 R$_1$=15.27 B$_2$=17.19 R$_2$=15.38 I=13.54}\\
\enddata
\tablenotetext{a}{The catalogs are the Two Micron All Sky Survey (2MASS, \citealp{cutri03}) and the United States Naval Observatory 
(USNO-B1.0 and USNO-A2.0, \citealp{monet98,monet03}). The astrometric accuracy of 2MASS and USNO catalogs are $\leq$0$\farcs$1 
\citep{skrut06} and 0$\farcs$2 respectively \citep{deu99,assafin01,monet03}.}
\tablenotetext{b}{The angular distance from the center of the {\emph{INTEGRAL}} error circle/The angular distance from the 
{\emph{Chandra}} position reported by T12}
\end{deluxetable*}
\begin{deluxetable*}{lcccccc}
\tablecolumns{7}
\tablewidth{0pc}
\tablecaption{Ew Measurements Of Candidate Counterparts To Igr J04069$+$5042 and Igr J06552$-$1146\label{MDMwidths}}
\tablehead{
\colhead{} & \colhead{} &\multicolumn{3}{c}{IGR J04069$+$5042} & \multicolumn{2}{c}{IGR J06552$-$1146}}
\startdata
& & J04064392$+$5044469 & &J04064872$+$5039316 & &J06545833$-$1149119\\ 
\cline{3-3} \cline{5-5} \cline{7-7}\\
SrII & $\lambda$4077 & \nodata & & 0.65$\pm$0.11 & & \nodata\\
H$\delta$ &  & \nodata & & 2.42$\pm$0.18 & & \nodata\\
CaI &$\lambda$4226 & \nodata & & 0.94$\pm$0.13 & & \nodata\\
H$\beta$ & & $\dots$ & &2.8$\pm$0.23 & \\
CaI triplet &$\lambda$6103 & 0.53$\pm$0.06 & & $<$0.07& &0.56$\pm$0.04\\
            &$\lambda$6122 &  1.00$\pm$0.04 & &$<$0.05& &0.98$\pm$0.14 \\
            &$\lambda$6162&  1.95$\pm$0.30 & &$<$0.3& &1.42$\pm$0.20\\
H$\alpha$ & & $-$1.07$\pm$0.04 & &2.41$\pm$0.08& &$-$0.45$\pm$0.05\\
\cline{2-7}\\
Spectral Class & & K5--7 Ve & &F5--8 III--V & & K5--8 Ve\\
\enddata
\tablecomments{EW measurements are given in \AA\ units and by convention, positive values denote the absorption lines.}
\end{deluxetable*}
\begin{deluxetable*}{lcccccccc}
\tablecolumns{9}
\tablewidth{0pc}
\tablecaption{Ew Measurements Of Counterparts To IGR J21188$+$4901 and IGR J22014$+$6034\label{TUGwidths}}
\tablehead{
\multicolumn{9}{c}{IGR J21188$+$4901}}\\
\startdata
& & & & J21182288$+$4906259 & &J21183906$+$4858049& & \\
\cline{5-5} \cline{7-7}\\
H$\alpha$ & & & &$-$1.9$\pm$0.24 & &$<$0.03 & &\\
\cline{2-9}\\
Spectral Class & & & & M3--5 Ve & & K7--M4 V & &\\
\cutinhead{IGR J22014$+$6034}
& &J22002116$+$6033420 & &J22010934$+$6034088 & &J22015416$+$6038094& &J22020837$+$6030425\\ 
\cline{3-3} \cline{5-5} \cline{7-7} \cline{9-9}\\
CaII H \& K &$\lambda$3934&$-$1.02$\pm$0.51& & \nodata & & \nodata & & \nodata\\
             &$\lambda$3968&$-$1.36$\pm$0.07& & \nodata & & \nodata & & \nodata\\
H$\beta$ & &\nodata& & 4.22$\pm$0.01 & &\nodata \tablenotemark{*}& &\nodata\\
FeI &$\lambda$5269 & \nodata& &1.07$\pm$0.01& & \nodata & &\nodata \\
H$\alpha$ & & $-$2.15$\pm$0.03 & &3.60$\pm$0.01 & &2.26$\pm$0.01& &$-$3.8$\pm$0.05\\
CaII triplet &$\lambda$8498& \nodata & &1.82$\pm$0.05 & &1.25$\pm$0.04 & & \nodata\\ 
             &$\lambda$8542&\nodata& &2.49$\pm$0.03 & &2.42$\pm$0.02 & & \nodata\\
             &$\lambda$8662&\nodata& &2.01$\pm$0.01 & &2.36$\pm$0.01 & & \nodata\\
\cline{2-9}\\
Spectral Class & & K7--M4 Ve & &F5--G5 III--V& &G III--IV & &M2--5 Ve \\
\enddata
\tablecomments{EW measurements are given in \AA\ units and by convention, positive values denote the absorption lines.}
\tablenotetext{*}{Since H$\beta$ absorption of J22015416$+$6038094 is highly blended, its EW measurement is not given in the table.}
\end{deluxetable*}
\clearpage
\begin{deluxetable*}{lccc}
\tablecolumns{4}
\tablewidth{0pc}
\tablecaption{Summary Table\label{summary}}
\tablehead{\multicolumn{4}{c}{\textbf{IGR J04069$+$5042}}\\
\colhead{$^{Candidate}\#$Chandra Name} & \colhead{Spectral Class} & \colhead{Type} & \colhead{Notes}}
\startdata
$^{1}$CXOU J040643.9$+$504446 & K5-7 Ve  &  Active star; RS CVn?   & H$\alpha$ emission, lower SAP\tablenotemark{a} \\
$^{2}$CXOU J040648.7$+$503931 & F5-8 III-V  &  Late-type star  & No emission lines, higher SAP \\
\cutinhead{\textbf{IGR J06552$-$1146}}
$^{1}$CXOU J065458.3$-$114911 & K5-8 Ve  & Active Star; RS Cvn?  & H$\alpha$ emission, higher SAP  \\
$^{2}$CXOU J065523.6$-$114601 & $-$ &  ? & No optical spectrum, intermediate SAP  \\
$^{3}$CXOU J065529.5$-$114900 &  $-$ & ? & No optical spectrum, low SAP, positive hardness \\
\cutinhead{\textbf{IGR J21188$+$4901}}
$^{1}$CXOU J211822.9$+$490627 & M3-5 Ve  & Active Star; Flare star?  &  H$\alpha$ emission, higher SAP  \\
$^{2}$CXOU J211839.1$+$485806 & K7-M4 V  & Late-type star  & lower SAP but not detected in 2-10 keV  \\
\cutinhead{\textbf{IGR J22014$+$6034}}
$^{1}$CXOU J220021.1$+$603342 & K7-M4 Ve  & Active star  &  H$\alpha$ emission, high SAP, detected in 2-10 keV \\         
$^{2}$CXOU J220109.3$+$603409 & F5-G5 III-V & Late-type star  & No features, high SAP \\          
$^{3}$CXOU J220154.1$+$603809 & G III-IV & Late-type star & No features, lower SAP \\
$^{4}$CXOU J220208.4$+$603042 & M2-5 Ve & Active Star; Flare star? &  H$\alpha$ emission, higher SAP \\
\enddata
\tablenotetext{a}{SAP: Spurious Association Probability}
\end{deluxetable*}
\clearpage
\begin{figure*}[!htb]
\begin{center}
\begin{tabular}{lr}
\\[5ex]
\includegraphics[scale=0.5]{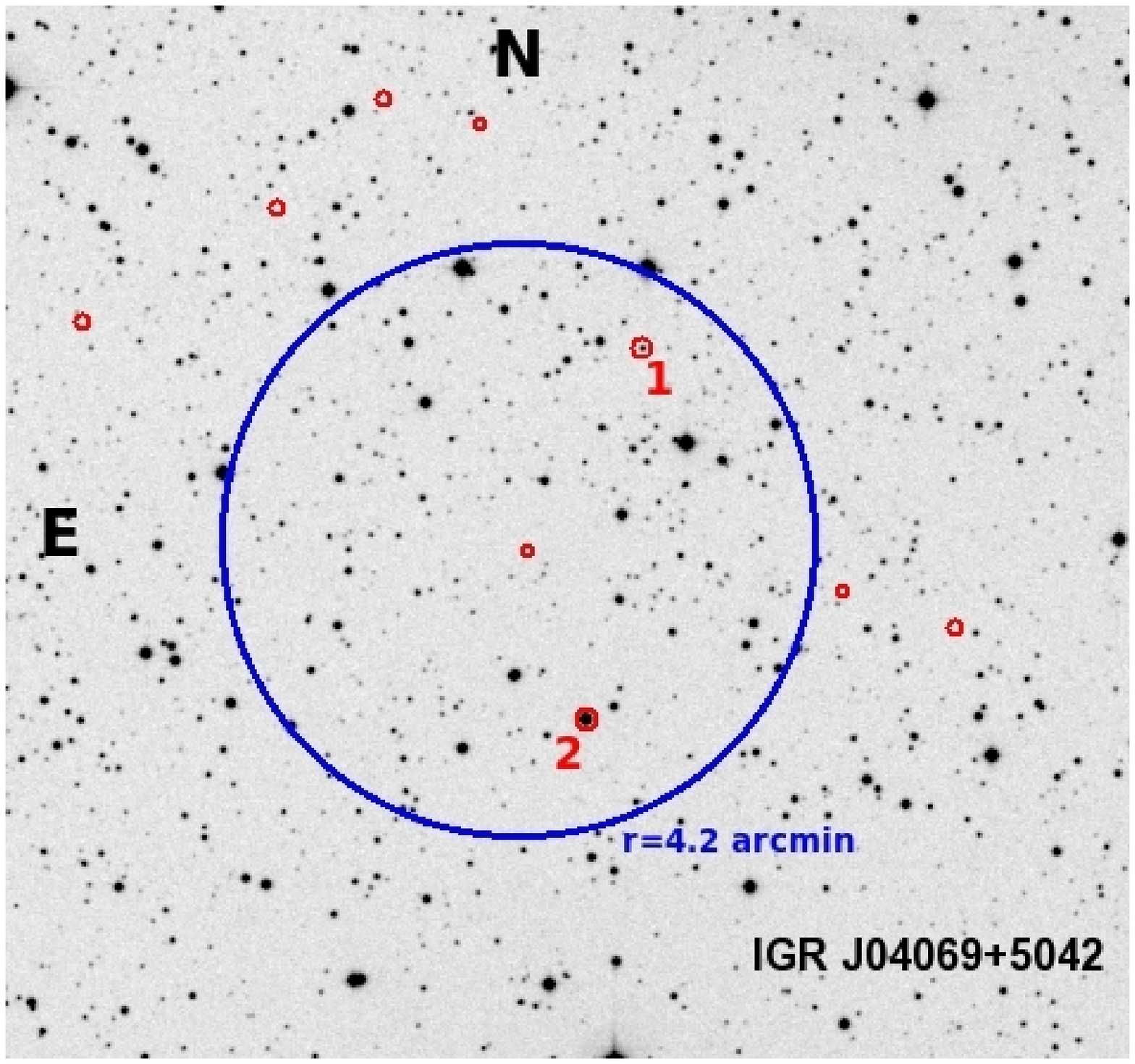} & \includegraphics[scale=0.5]{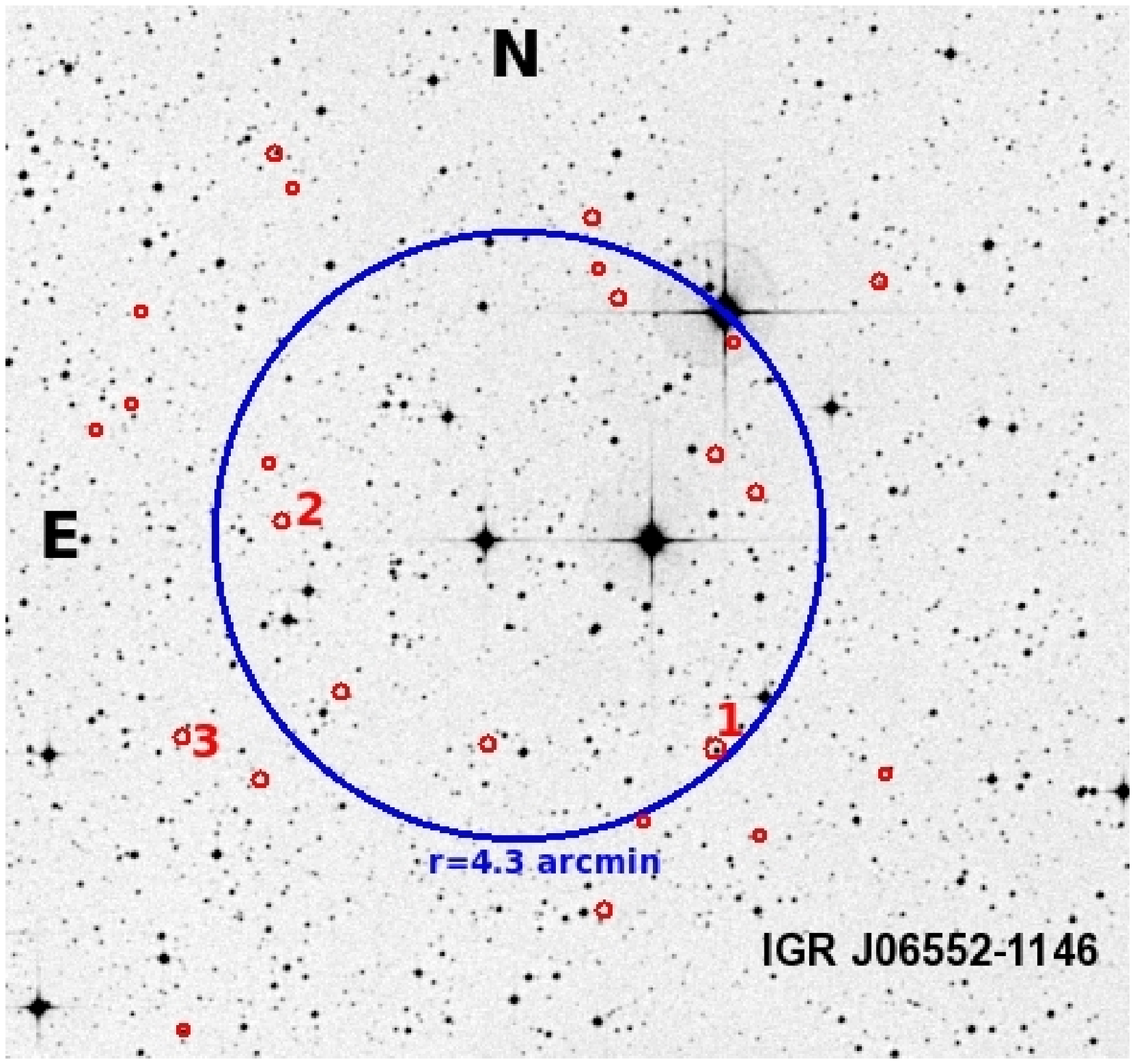}\\
\\[0.01ex]
\includegraphics[scale=0.5]{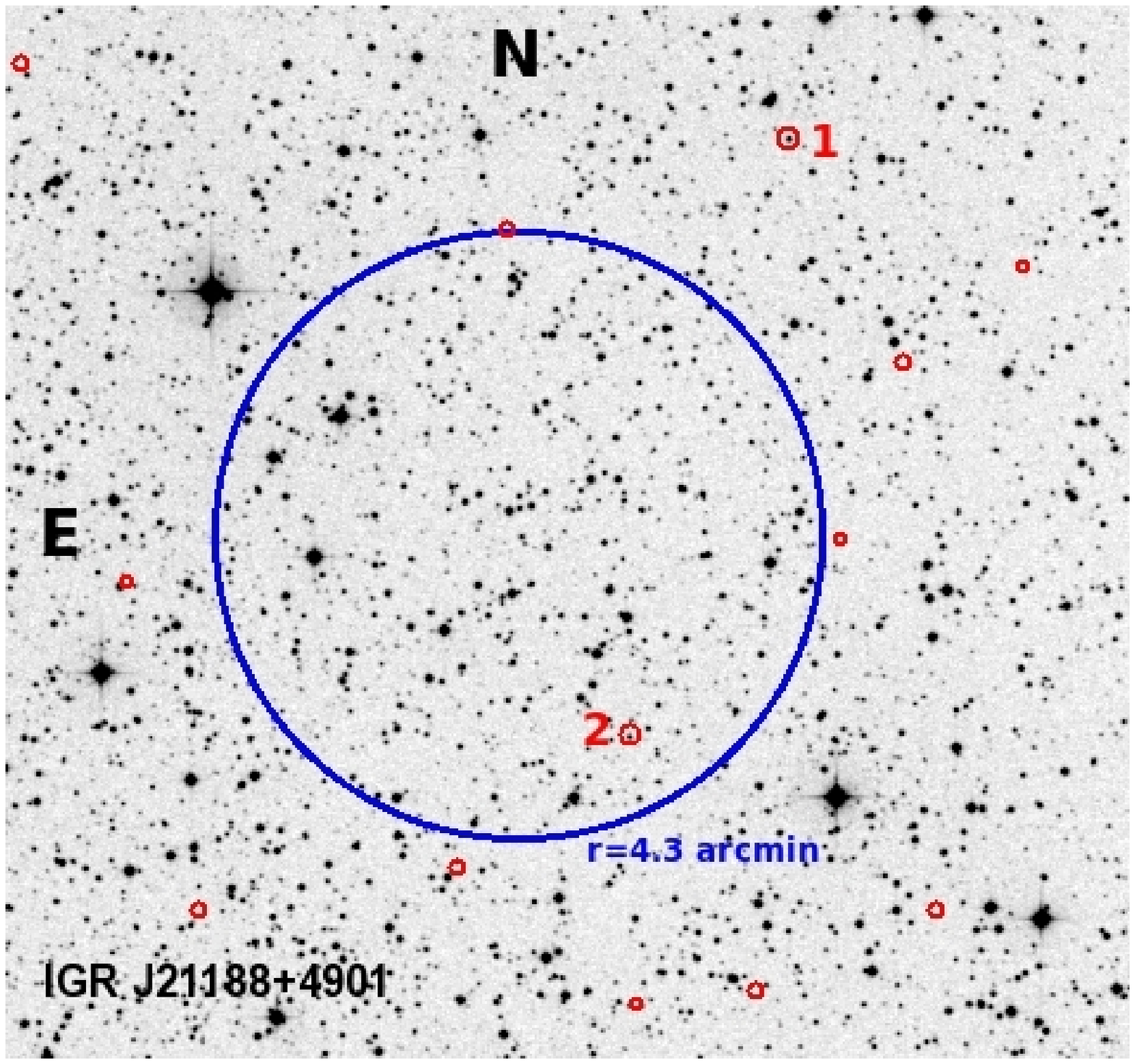} &\includegraphics[scale=0.5]{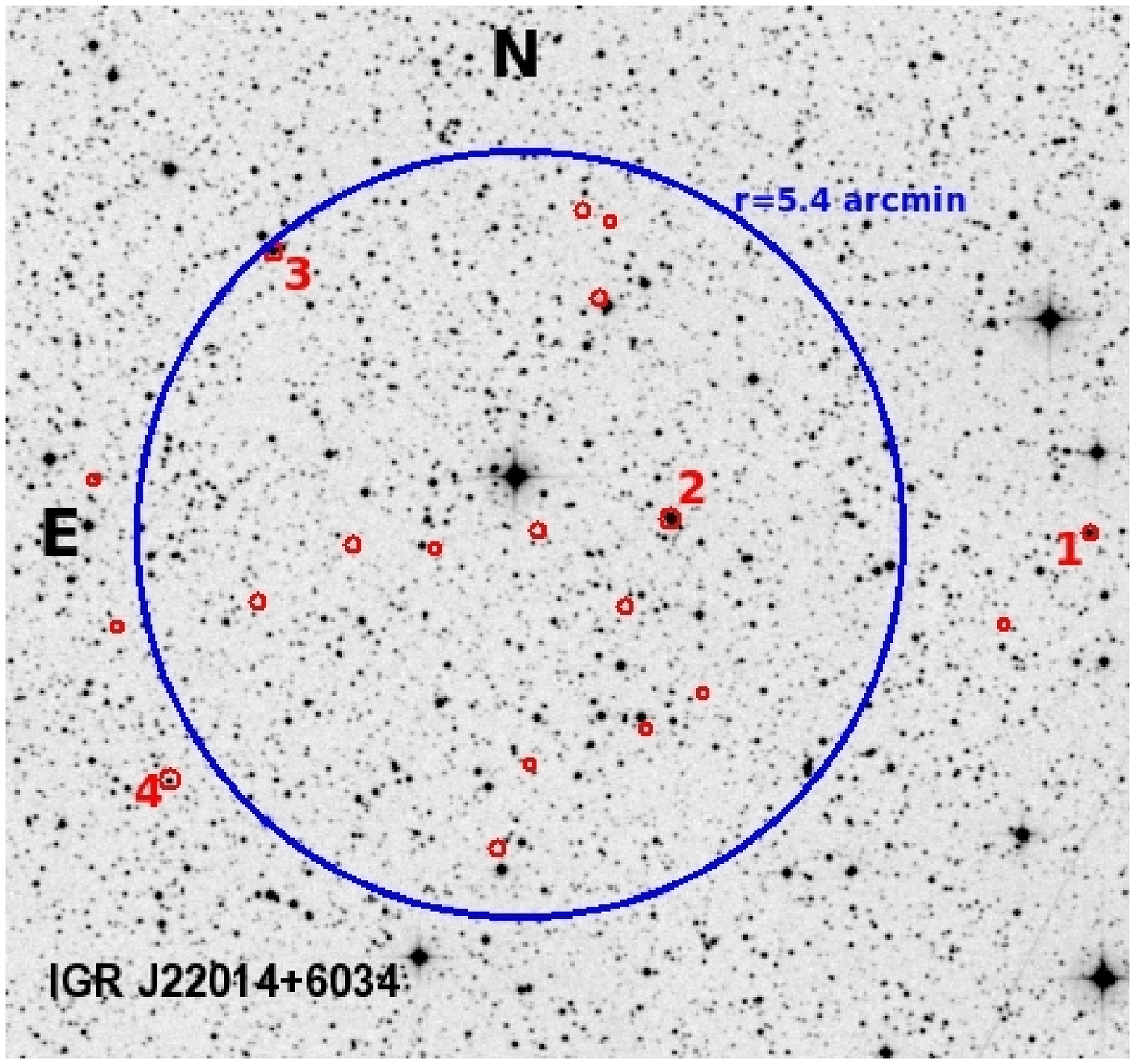}\\
\end{tabular}
\\[8.5ex]
\caption{DSS-R images of the fields around IGR J04069$+$5042 (upper left panel), IGR J06552$-$1146 (upper right panel), 
         IGR J21188$+$4901 (lower left panel) and IGR J22014$+$6034 (lower right panel).  
         Blue circles indicate the 90\% confidence radii of {\emph{INTEGRAL}} error circle, whereas red circles show 
         the positions of the candidate {\emph{Chandra}} counterparts in these fields given in T12. 
         For each frame the observed candidates which are presented in this work are labeled with numbers.}
\label{DSS}
\end{center}
\end{figure*}
\begin{figure*}[!ht]
 \centering
\begin{tabular}{c} 
\\[5ex]
\includegraphics[angle=-90,scale=0.65]{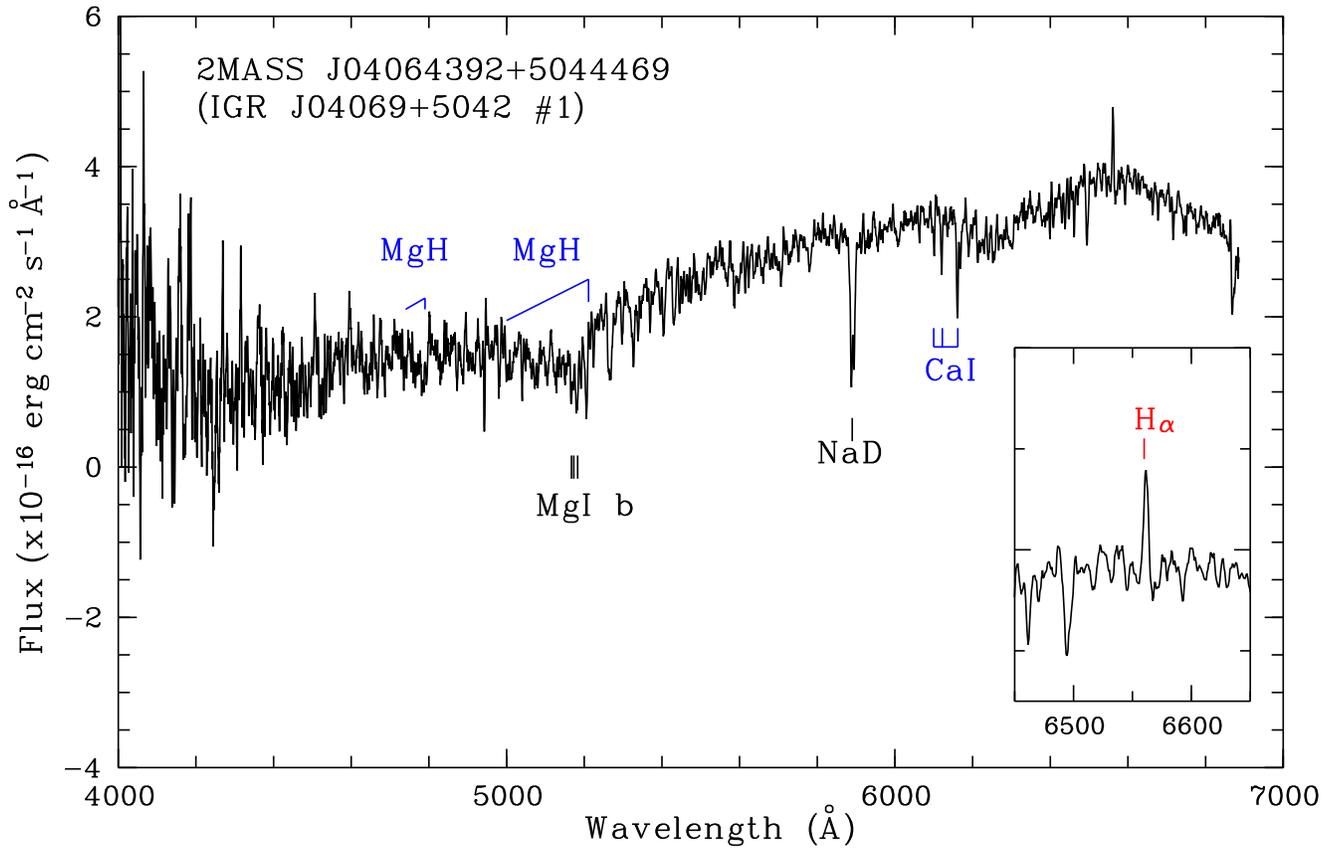}
\end{tabular}
\\[8.5ex]
\caption{The spectrum of candidate \#1 of \I04069$+$5042. A weak H$\alpha$ emission 
line can be distinguished from the noise level. This source is identified as a K5-7 type main sequence star due to the existence of 
molecular MgH bands and the similarities between the stars of the same spectral type.}
\label{04069s}
\end{figure*}
\begin{figure*}[!ht]
 \centering
\begin{tabular}{c}
 \\[5ex]
 \includegraphics[angle=-90,scale=0.65]{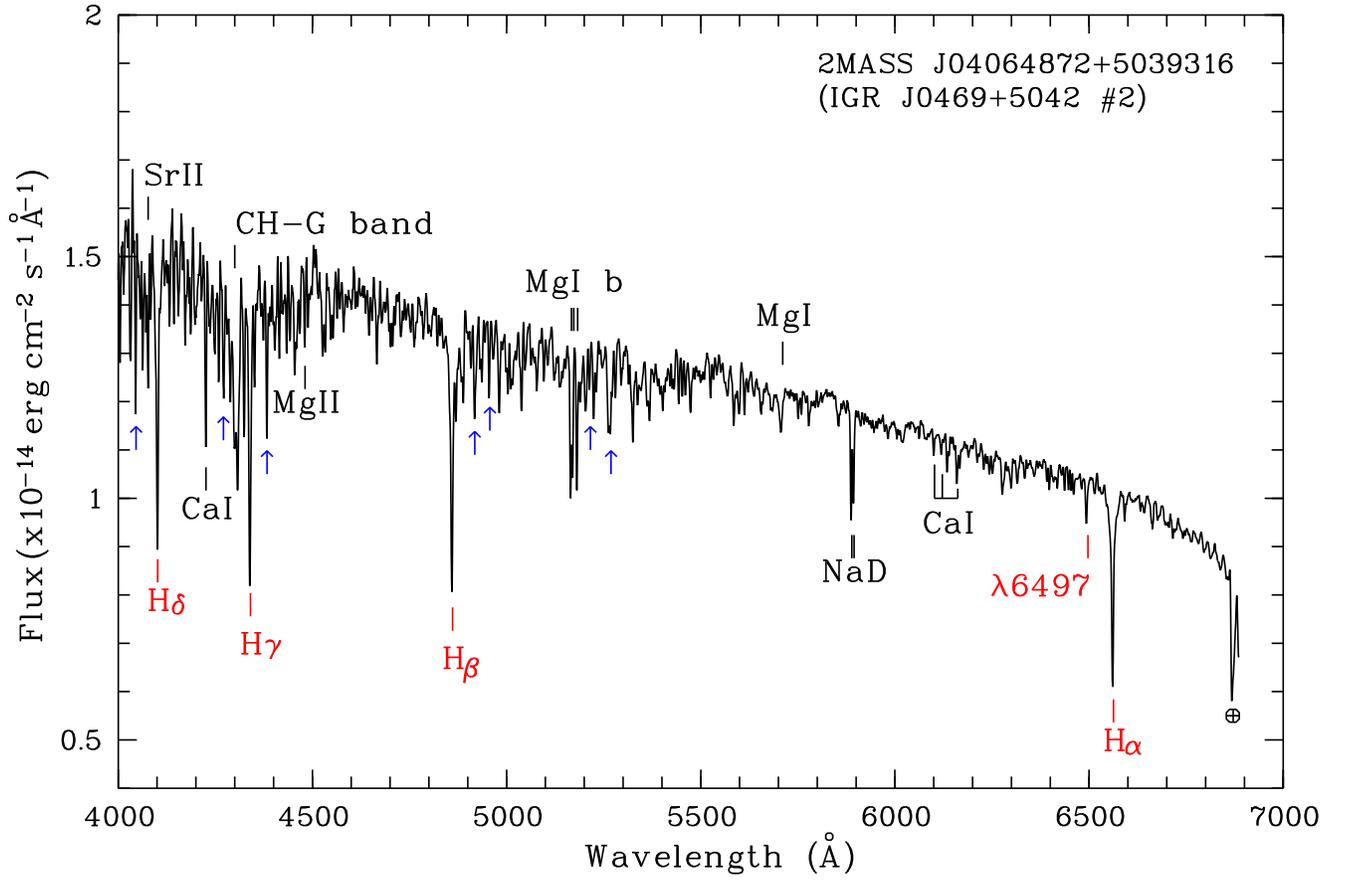}\\
\end{tabular}
\\[8.5ex]
\caption{The spectrum of candidate \#2 of \I04069$+$5042 shows the characteristics of an F type star. Blue arrows indicate the FeI 
lines at $\lambda$$\lambda$4046, 4271, 4383, 4819, 4957, 5216 and 5270 respectively whereas symbol $\oplus$ shows the telluric 
absorption band. The Balmer series lines and the CH-G band are the most prominent features. Neutral and singly ionized Magnesium lines, 
CaI triplet, NaD absorption and the blend of several metal lines at 6497 \AA\ are also visible.}
\label{5039316}
\end{figure*}
\clearpage
\begin{figure*}[!htb]
\begin{center}
\begin{tabular}{lr}
\\[20ex]
\includegraphics[scale=0.45]{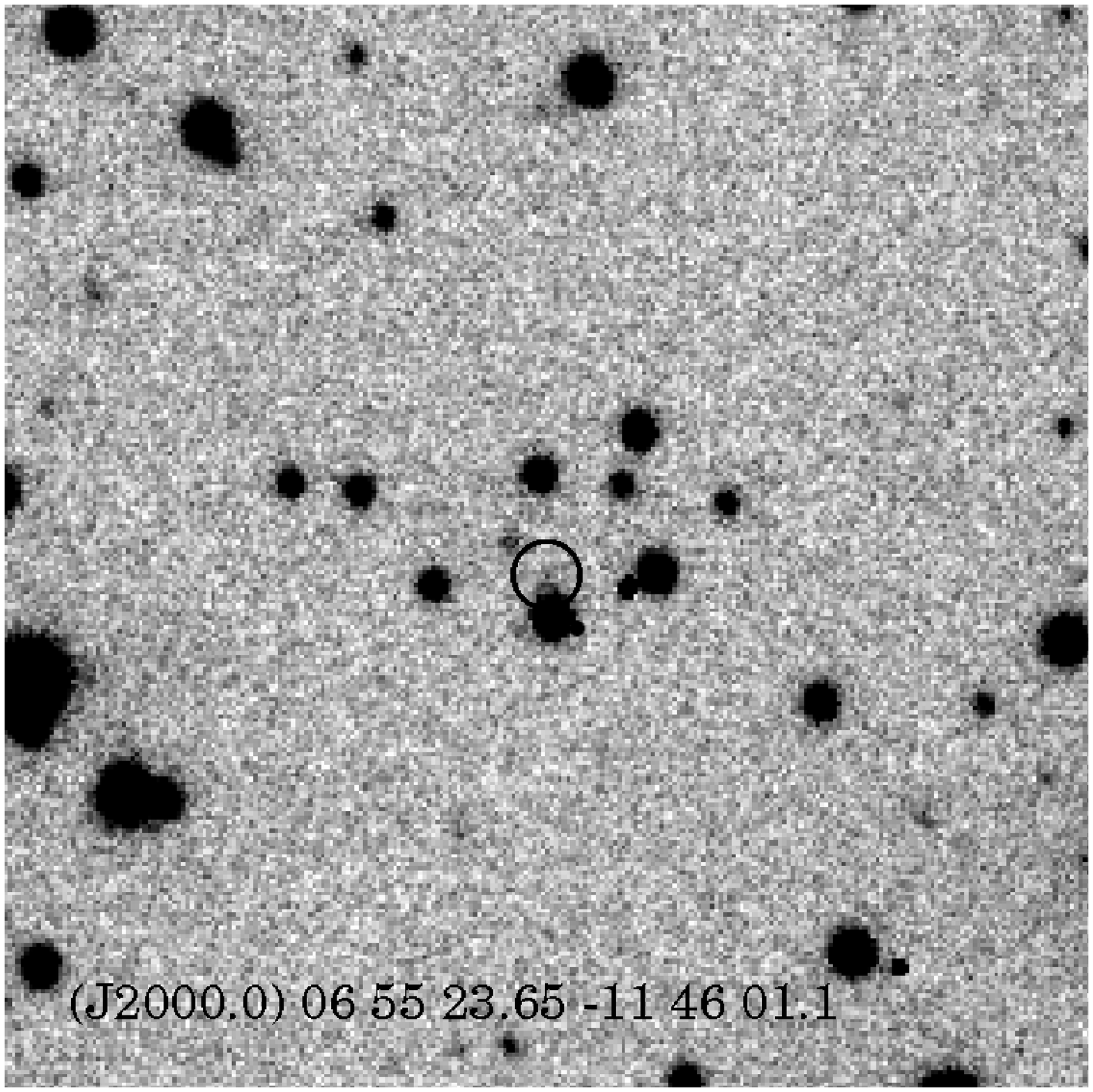} & \includegraphics[scale=0.45]{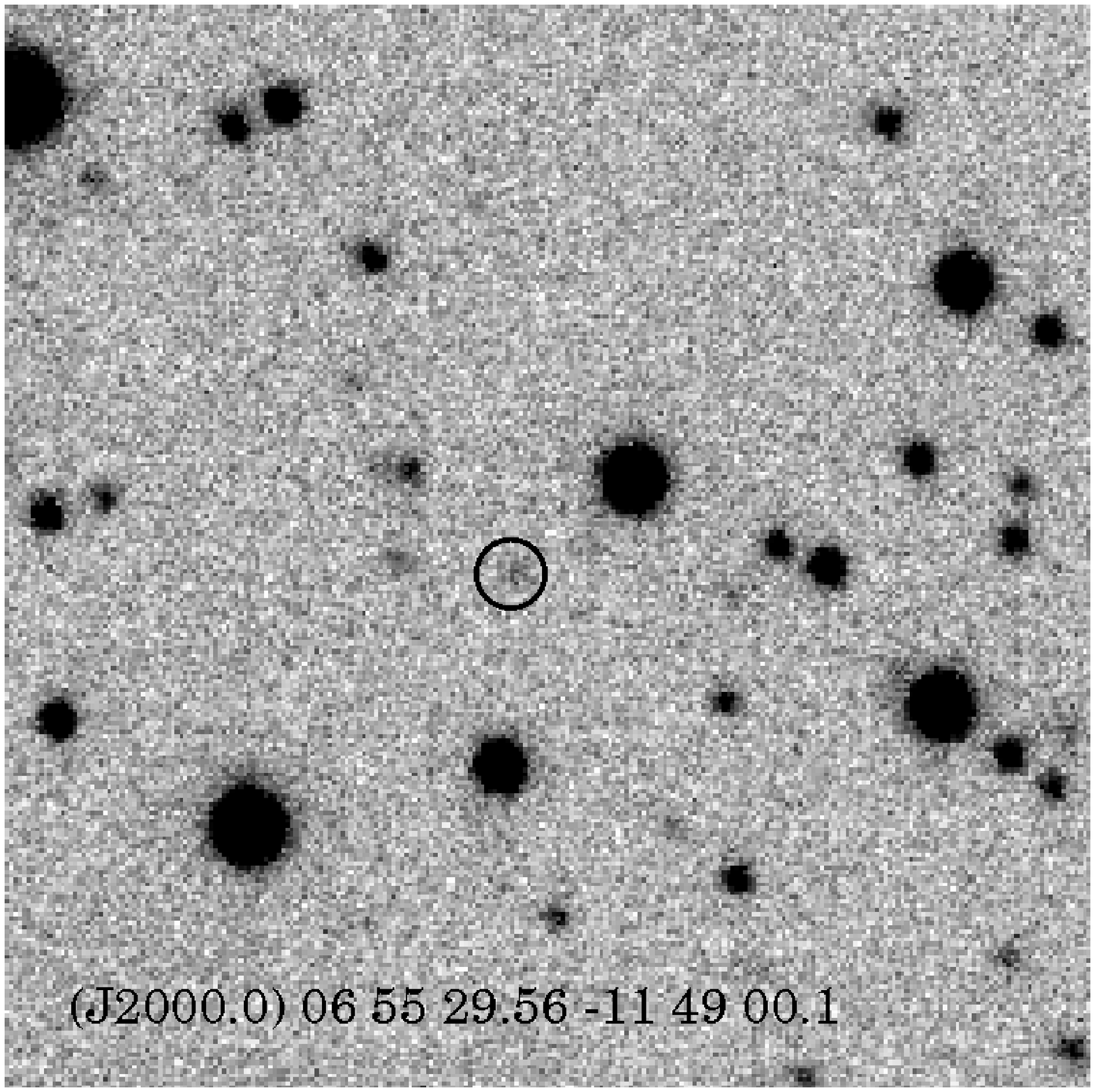}\\
\end{tabular}
\\[8.5ex]
\caption{$R$-band images of the field around the candidates \#2 (left panel) and \#3 (right panel) of \I06552$-$1146. The black circles 
show the positions of \emph{Chandra} sources, but the radii are larger than the error radii for clarity. The \emph{Chandra} error radii 
are 0\farcs68 and 0\farcs81 respectively. For the candidate \#2, the source in the error circle is ambiguous while the \emph{Chandra} 
position is not coincident with any optical or IR counterpart in the catalogs for the latter. The $R$-band magnitude of the weak source 
in the error circle for the candidate \#3 is found to be 22.17$\pm$0.14 from the aperture photometry.}
\label{MDM06}
\end{center}
\end{figure*}
\clearpage
\begin{figure*}[!ht]
\centering
\begin{tabular}{c}
\\[5ex]
\includegraphics[angle=-90,scale=0.65]{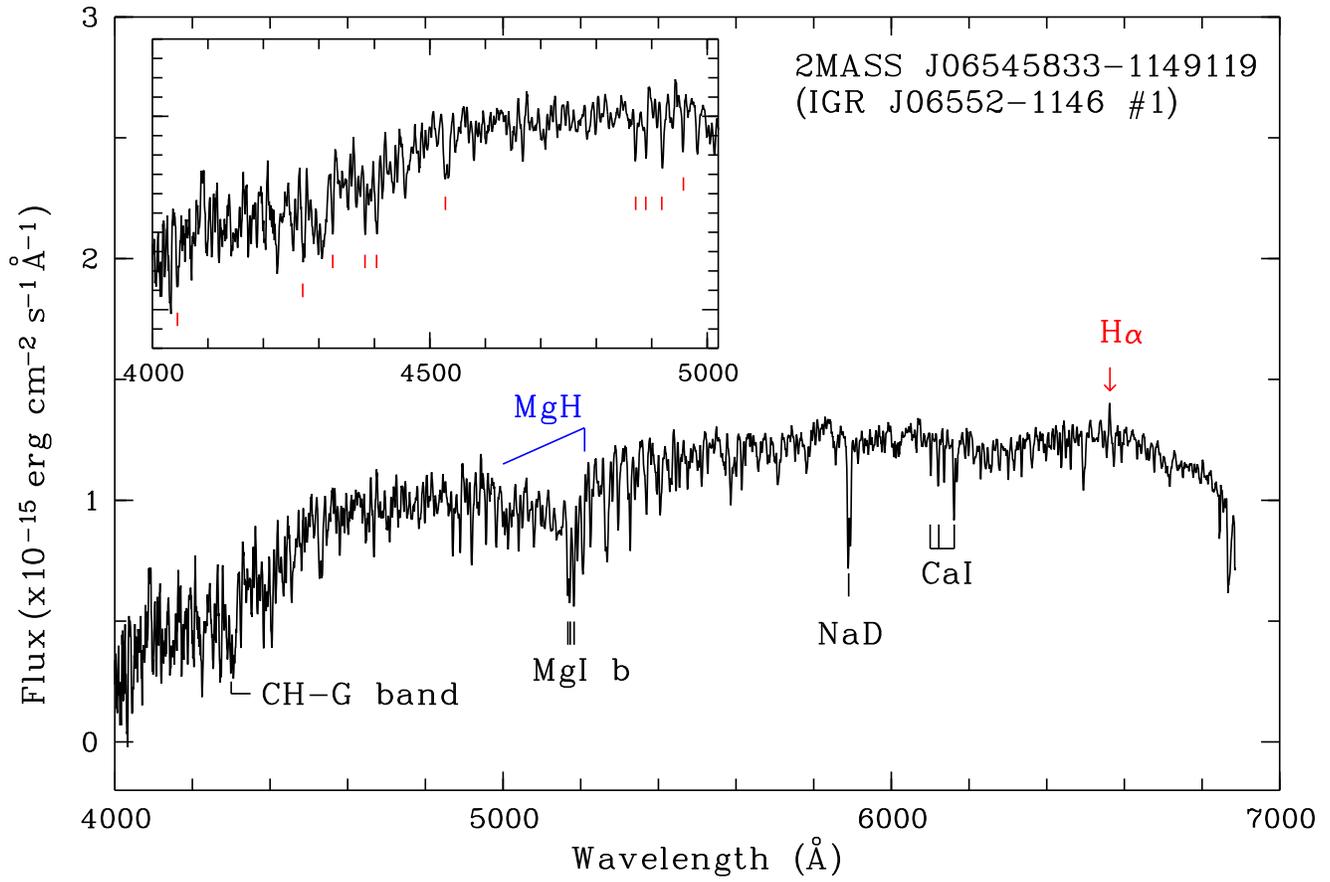}\\
\end{tabular}
\\[8.5ex]
\caption{The flux-calibrated spectrum of candidate \#1 to \I06552$-$1146. FeI lines (red bars) at $\lambda$$\lambda$4046, 4271, 4325, 
4383, 4404, 4528, 4871, 4889, 4918, 4957 are shown in a separate panel for clarity. A weak H$\alpha$ emission is also seen in the main panel.}
\label{06552s}
\end{figure*}
\clearpage
\begin{figure*}[!ht]
\begin{center} 
\begin{tabular}{c}
\\[5ex]
\includegraphics[angle=-90,scale=0.65]{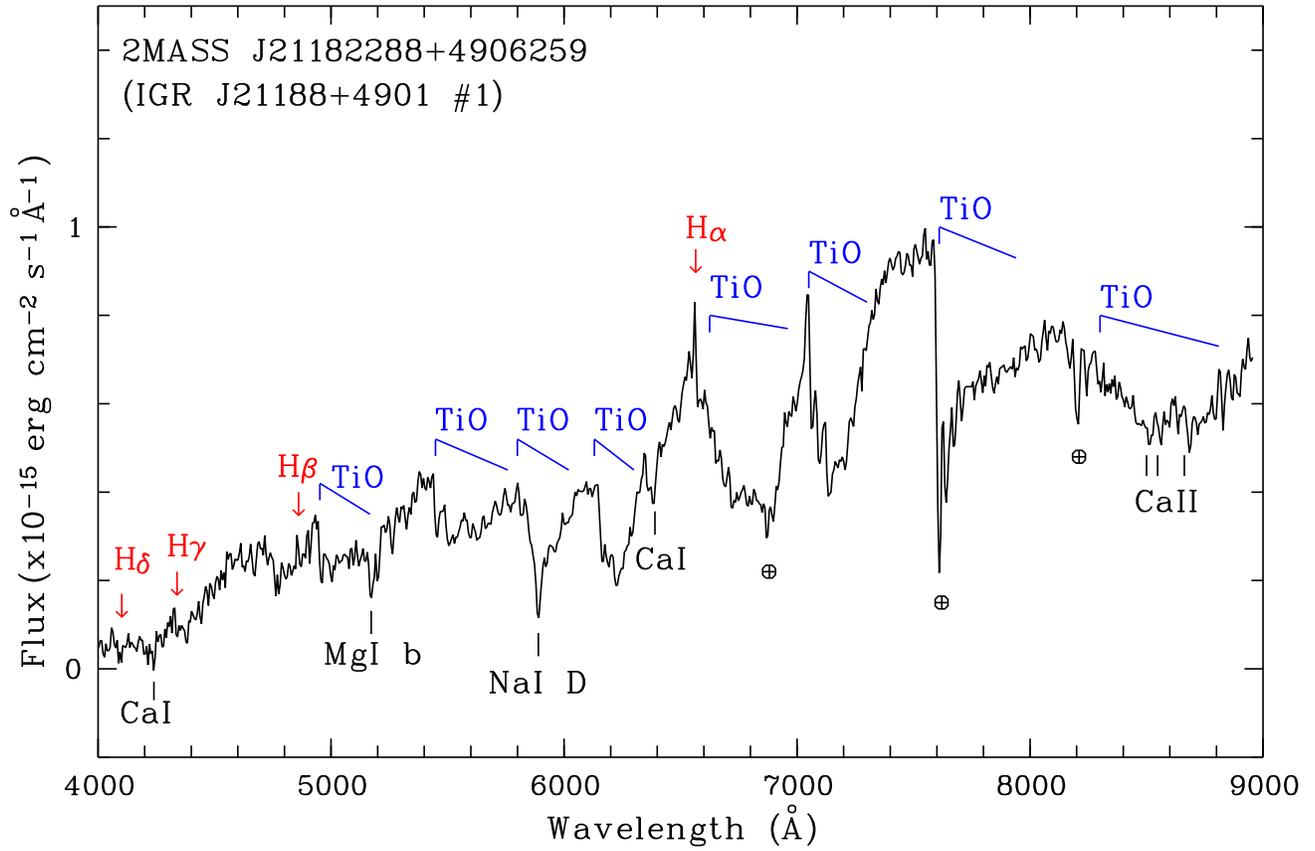}\\
\end{tabular}
\\[8.5ex]
\caption{The identified features in flux-calibrated spectrum of candidate \#1 to \I21188$+$4901. The spectrum contains both broad 
molecular TiO bands with a red continuum and a prominent H$\alpha$ emission, indicating an M type chromospherically active star. The 
telluric absorption bands are denoted by $\oplus$ symbol.}
\label{21188n1}
\end{center}
\end{figure*}
\clearpage
\begin{figure*}[!ht]
 \begin{center}
\begin{tabular}{c} 
\\[5ex]
\includegraphics[angle=-90,scale=0.65]{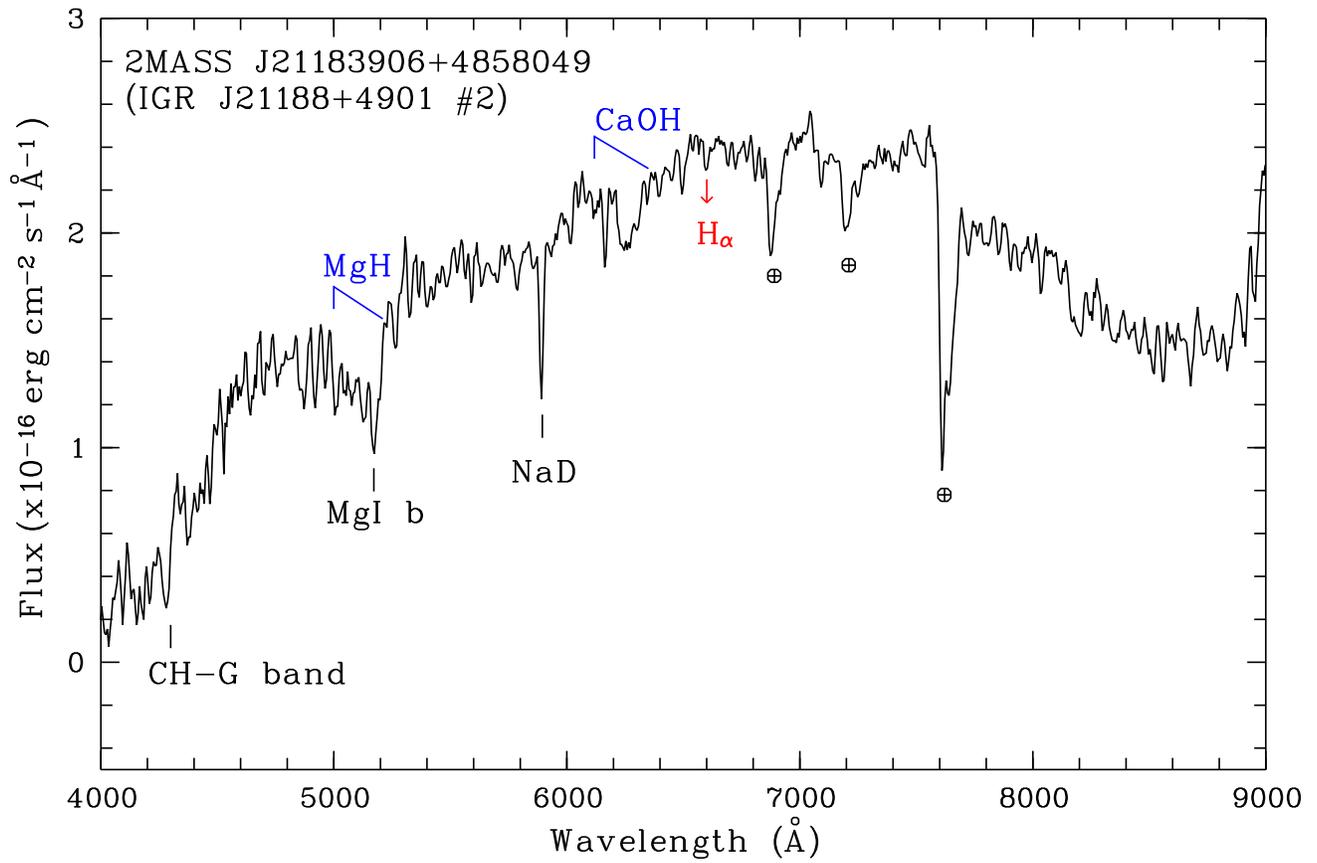}\\
\end{tabular}
\\[8.5ex]
\caption{The flux-calibrated spectrum of candidate \#2 to \I21188$+$4901. The source has typical features of a late-type star. 
The symbol $\oplus$ indicates the telluric absorption bands.}
\label{21188n2}
\end{center}
\end{figure*}
\clearpage
\begin{figure*}[!ht]
 \centering
\begin{tabular}{c}
\\[5ex]
\includegraphics[angle=-90,scale=0.65]{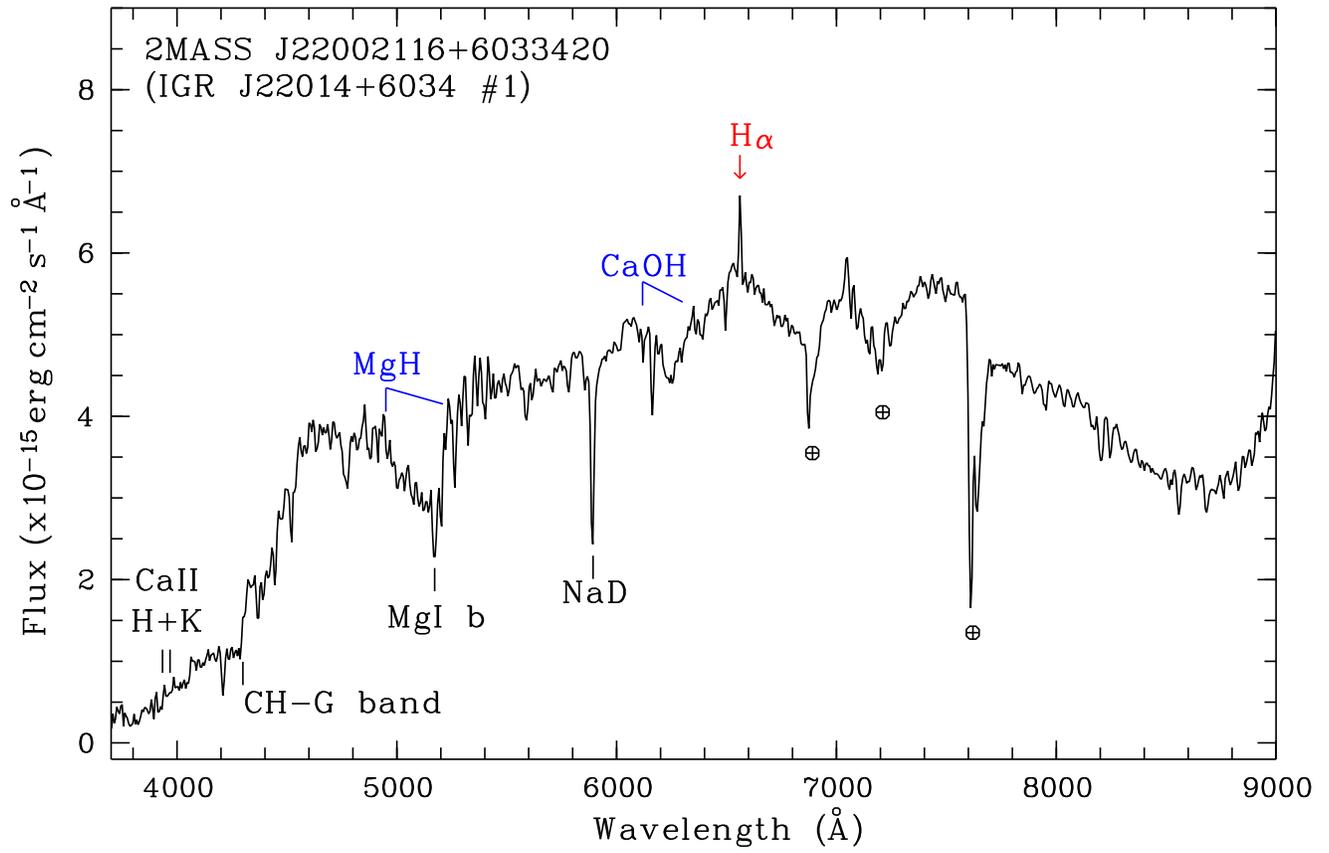}\\
\end{tabular}
\\[8.5ex]
\caption{The flux-calibrated spectrum of candidate \#1 to \I22014$+$6034. The emission lines of CaII H \& K and H$\alpha$ 
can be considered as indicators of a chromospherically active star.}
\label{22014n1}
\end{figure*}
\begin{figure*}[!ht]
\centering
\begin{tabular}{c}
\\[5ex]
\includegraphics[angle=-90,scale=0.65]{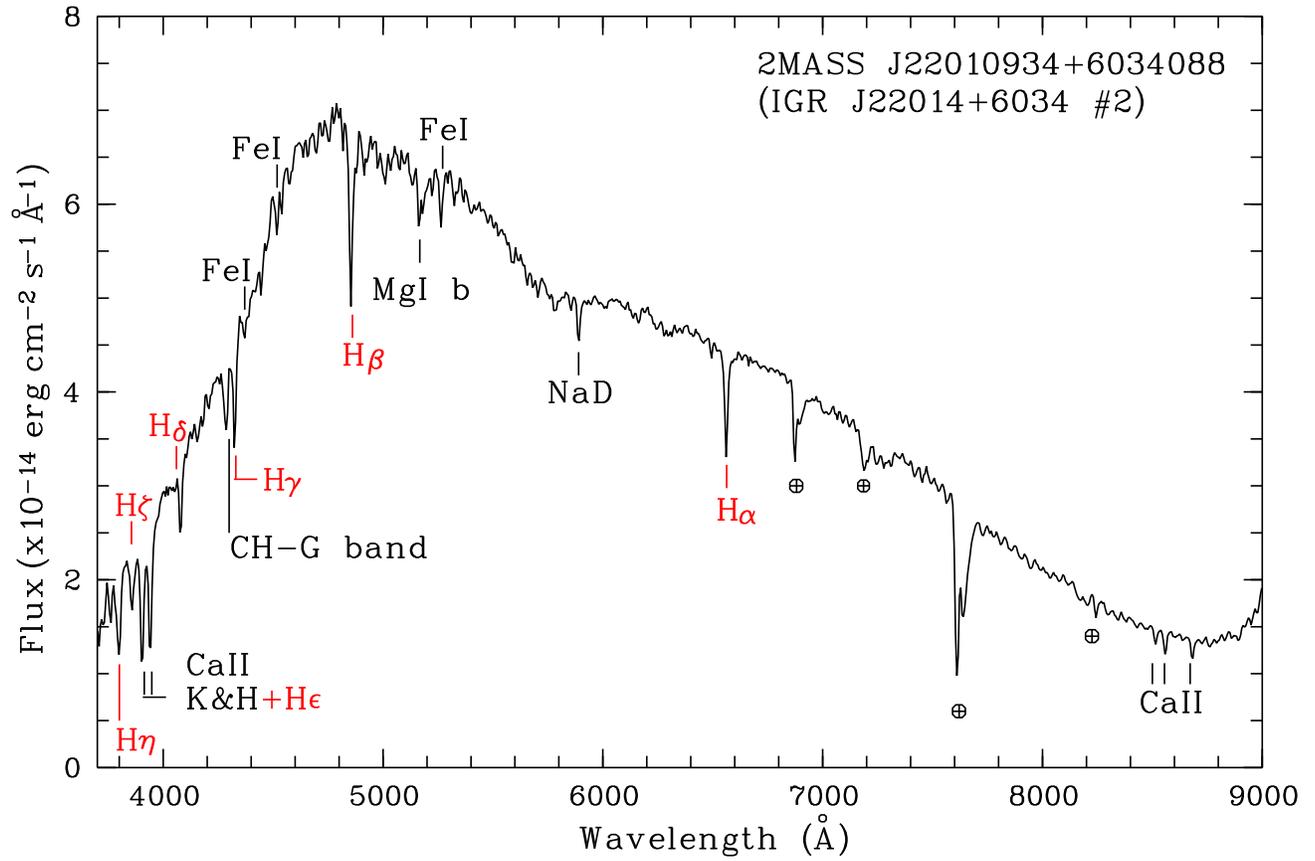}\\
\end{tabular}
\\[8.5ex]
\caption{The flux-calibrated spectra of 2MASS J22010934$+$6034088, candidate \#2 to \I22014$+$6034. The spectrum is dominated by 
Balmer series lines and metallic lines of FeI, MgI. The symbol $\oplus$ denotes the atmospheric telluric bands.}
\label{22014_n2}
\end{figure*}
\clearpage
%
\begin{figure*}[!ht]
\centering
\begin{tabular}{c}
\\[5ex]
\includegraphics[angle=-90,scale=0.65]{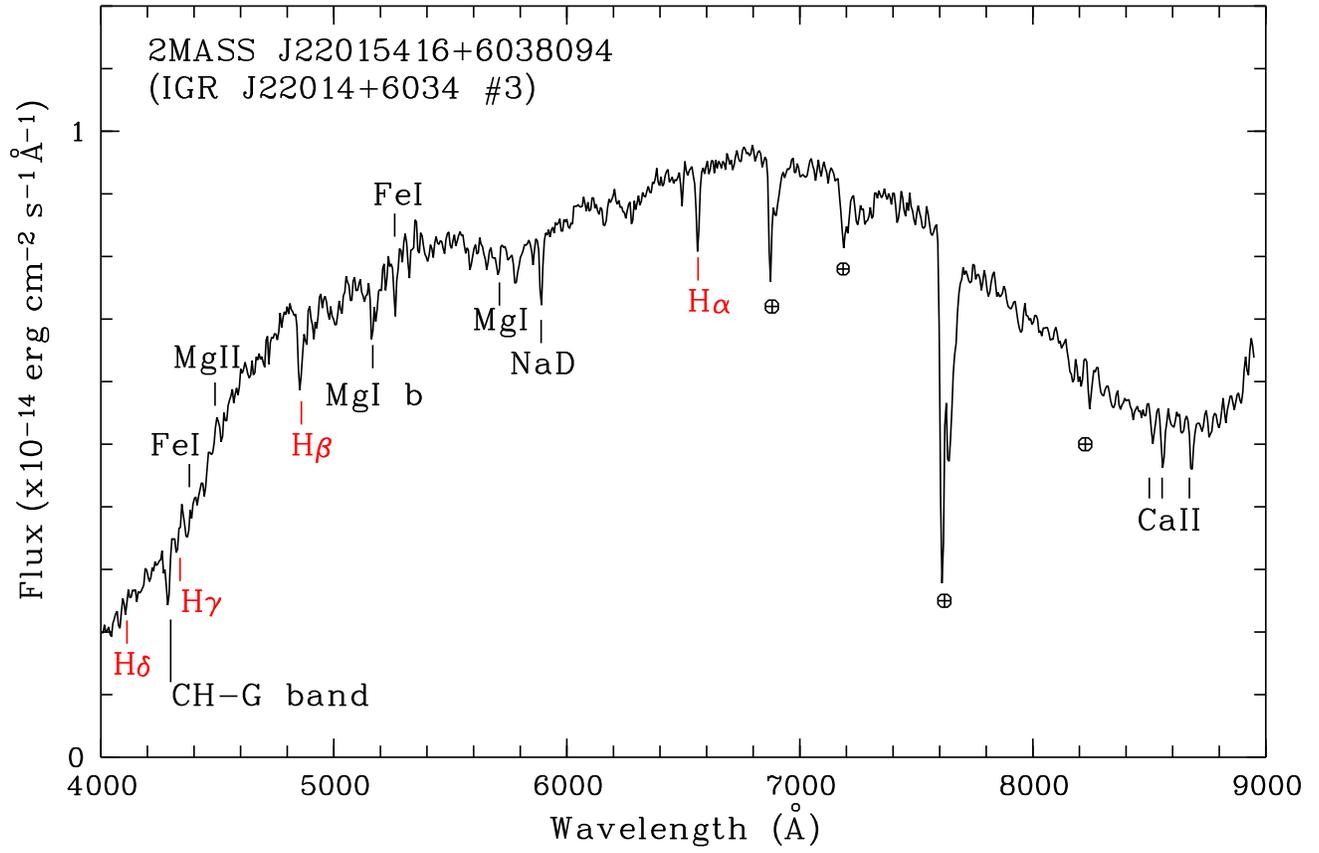}\\
\end{tabular}
\\[8.5ex]
\caption{The flux-calibrated spectrum of 2MASS J22015416$+$6038094, candidate \#3 to \I22014$+$6034. The main spectral features 
MgI, MgII and FeI lines in addition to Balmer series lines are labeled.The symbol $\oplus$ denotes the atmospheric telluric bands. 
This candidate is identified as a G type star.}
\label{22014_n3}
\end{figure*}
%
%
\clearpage
\begin{figure*}[!ht]
\centering
\begin{tabular}{c}
\\[5ex]
\includegraphics[angle=-90,scale=0.65]{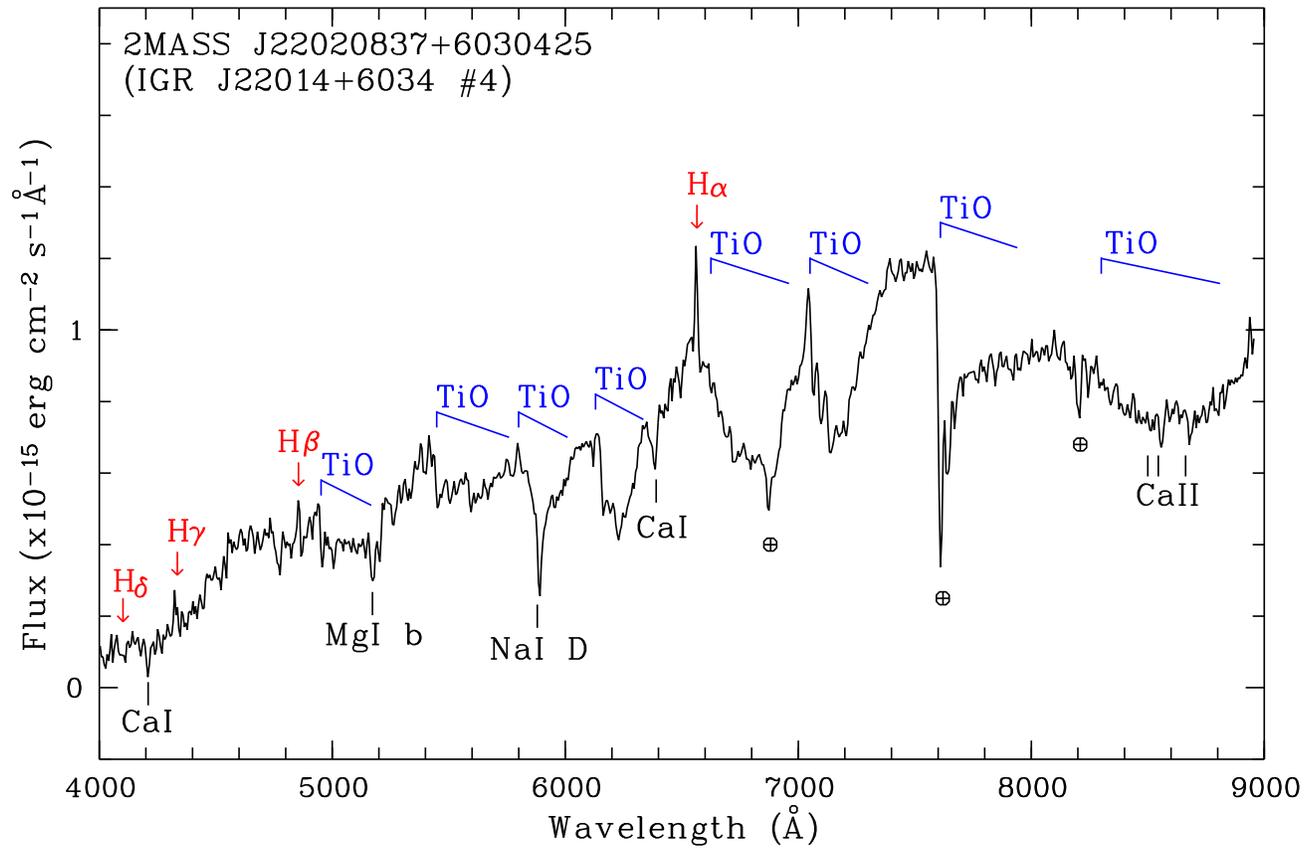}\\
\end{tabular}
\\[8.5ex]
\caption{The flux-calibrated spectrum of candidate \#4 to \I22014$+$6034. The spectrum shows typical spectral features of a M 
type main sequence star dominated by strong molecular bands of TiO.}
\label{22014_n4}
\end{figure*}
\begin{figure*}[!ht]
 \centering
 \includegraphics[]{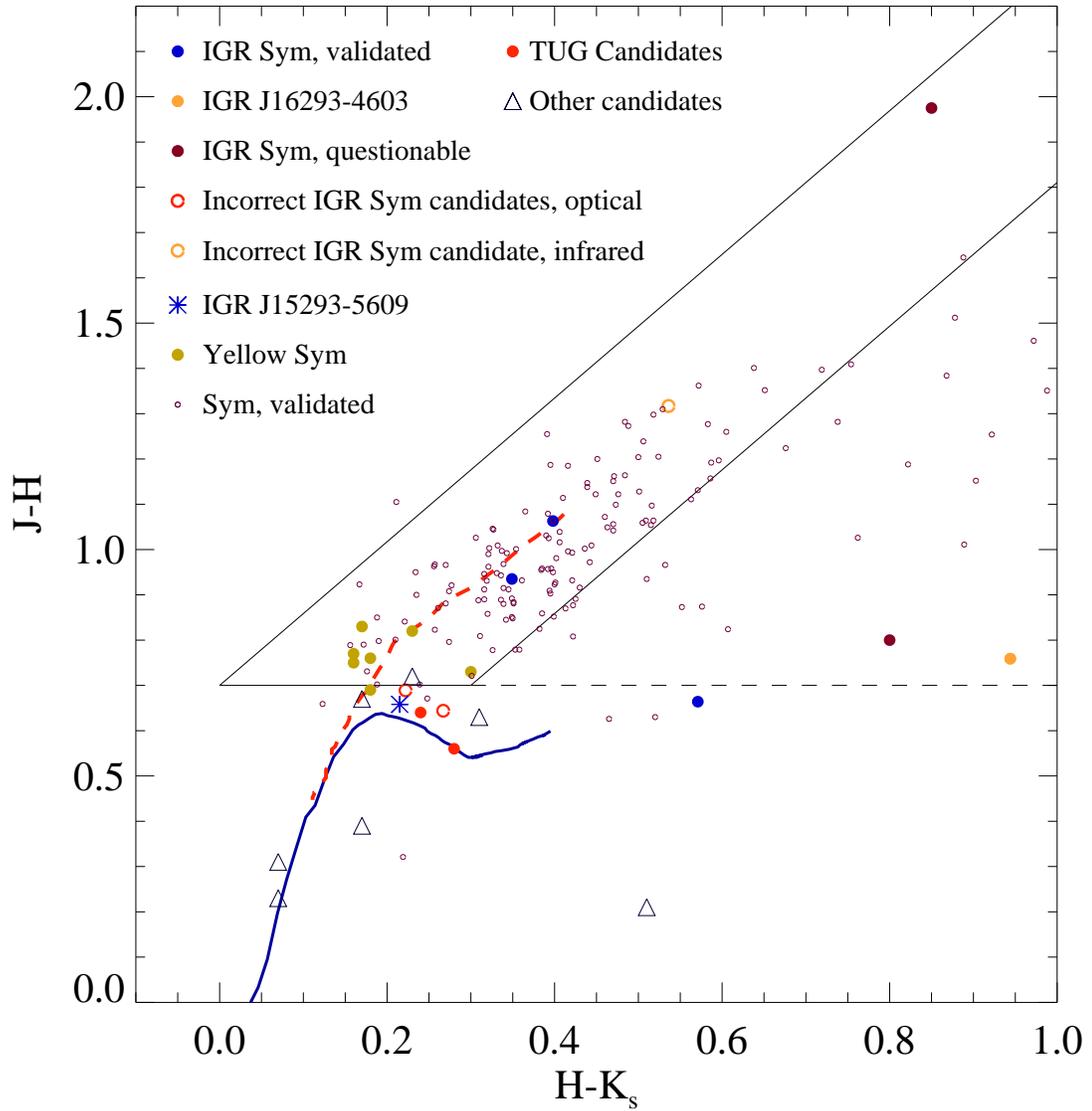}
\caption{$H-Ks$ vs $J-H$ diagram for all sources discussed in this work. The validated symbiotics are taken from 
\citet{phil07}, and shown with small, empty, purple circles. The solid lines enclose the S type symbiotics, 
whereas the single dashed line and the first solid line enclose D type symbiotics \citep{Corradi08}. The locus of points for main 
sequence stars and red giant branch stars are represented by a blue solid line and a red dashed line respectively. The questionable 
symbiotic counterparts (filled brown symbols) are for IGR J16358$-$4726 and IGR J17497$-$2821, the incorrect IGR symbiotic candidates 
(empty red circles) are for IGR J11098$-$6457 and IGR J17197$-$3010, and the incorrect IGR symbiotic identified through $K$-band 
spectroscopy is for IGR J16393$-$4643. See \S\ref{subsec:symbio} for more details.}
\label{symbio}
\end{figure*}
\end{document}